\documentclass[journal]{IEEEtran}
\IEEEoverridecommandlockouts
\usepackage{cite}
\usepackage{amsmath,amssymb,amsfonts}
\usepackage{algorithm}
\usepackage{algorithmic}
\usepackage{graphicx}
\usepackage{textcomp}
\usepackage{xcolor}
\usepackage{amsmath}
\usepackage{multirow}
\usepackage{graphicx}
\usepackage{tabularx}
\usepackage{amsthm}
\usepackage{subfigure}
\usepackage{booktabs}

\usepackage{subfig}

\DeclareMathOperator{\diag}{diag}

\DeclareMathOperator{\sinc}{sinc}

\begin{document}

\title{BER-Aware Joint Parameter Optimization of Rydberg Atomic Quantum Receivers}
\author{A. K. Mandal and M. Shen
\thanks{A. K. Mandal and M. Shen are affiliated with the Department of Electronics System at Aalborg University, Denmark.}
}
\maketitle
%
\begin{abstract}
Rydberg-atom quantum receivers (RAQRs) offer highly sensitive RF reception, but their communication performance depends jointly on atomic response, optical transduction, receiver noise, bandwidth, and wireless-channel conditions. This paper develops a BER-aware communication-theoretic framework for superheterodyne RAQR-assisted wireless communication, linking four-level Rydberg-atom dynamics and coherent optical detection to multipath propagation and digital communication performance. Closed-form instantaneous and average SNR and BER expressions are derived for square Gray-coded $M$-QAM over correlated Rayleigh fading. The RAQR operating point is then jointly optimized over atomic, optical, RF, and geometric parameters using analytical block-wise updates within a fixed-point procedure subject to physical and bandwidth constraints. Numerical results demonstrate substantial gains in low-power reception, bandwidth utilization, robustness to detuning, fading and delay spread, outage, and communication coverage compared with conventional, advanced, and fixed-point RAQR receivers. The results demonstrate that BER-aware joint optimization translates the intrinsic sensitivity of Rydberg-atom sensing into tangible communication-level gains.
\end{abstract}
%
%
\section{Introduction}
Evolution of wireless systems toward higher carrier frequencies, wider bandwidths, heterogeneous spectrum access, and compact hardware increasingly challenges conventional receivers, which face tradeoffs among sensitivity, bandwidth, interference, frequency agility, and hardware complexity. Rydberg-atom quantum receivers (RAQRs) offer an alternative by directly sensing RF fields through highly sensitive Rydberg transitions and optically interrogating the resulting atomic response \cite{Holloway2014,Fancher2021,Cox2018}. In a ladder-type electromagnetically induced transparency (EIT), the RF field couples adjacent Rydberg states and modifies the transmitted probe amplitude, phase, or spectrum, enabling direct RF-to-optical transduction \cite{Sedlacek2012,Holloway2014}.

\subsection{State-of-the-Art and Motivation}
RAQR research has progressed from RF electrometry and traceable field measurement to signal reception, including AM, FM, phase, and high-order QAM \cite{Cox2018,anderson2020atomic,yuan2023rydberg,Cai2023}. Atomic superheterodyne reception further introduces an RF local oscillator (LO) to generate a coherent atomic beat response, enabling sensitive phase and frequency-resolved detection and a natural coherent-receiver architecture \cite{Jing2020,Cai2023}. Recent studies have consequently developed end-to-end models linking RF propagation, atomic transduction, optical detection, and receiver noise, demonstrating substantial communication-level SNR advantages \cite{gong2025rydberg,gong2026rydberg}. However, existing analyses generally assume fixed physical parameters or optimize only selected receiver parameters \cite{gong2025rydberg,gong2026rydberg}.

This limitation is significant because RAQR performance is jointly governed by atomic susceptibility, optical powers, RF-LO operating point, Rabi frequencies, detunings, atomic density, cell geometry, and EIT linewidth, while quantum projection, photon-shot, and electronic thermal noise respond differently to these parameters \cite{Holloway2014,Cox2018,gong2026rydberg}. Moreover, EIT bandwidth introduces a fundamental communication tradeoff: a narrow response suppresses noise but can distort the desired waveform, whereas a broad response admits more noise and modifies the effective multipath response. In our model, the effective number of channel taps depends on the minimum of the communication and RAQR bandwidths, directly coupling receiver physics to frequency-selective propagation.

The RF and optical parameters are strongly coupled: RF Rabi frequencies affect susceptibility and EIT linewidth, probe and LO fields determine coherent-detection gain and shot noise, detunings affect absorption and dispersion, and atomic density and cell length determine optical depth \cite{Holloway2014,Jing2020,Cai2023,sandidge2024resonant,wu2024enhancing}. Together with weak-probe, two-photon-resonance, bandwidth-matching, low-IF, and phase constraints, these couplings make communication-level optimization nonlinear and challenging; changes in the effective channel-tap count can further render the BER objective piecewise continuous.

These observations motivate a BER-aware joint optimization framework that treats the RAQR operating point as an integral part of wireless receiver design rather than as a fixed front end. We develop an end-to-end analytical model connecting four-level atomic dynamics, coherent optical detection, multipath propagation, receiver noise, and BER, and use it to jointly optimize atomic, optical, RF, and communication parameters under physical constraints. This approach is timely as Rydberg quantum radios expand toward SISO/MIMO, sensing, multiband, and next-generation wireless systems \cite{gong2025rydberg,gong2026rydberg}, while practical deployment remains limited by bandwidth, noise, environmental broadening, laser stability, and integration \cite{Fancher2021,Borowka2024,Babusenan2026}. The central premise is that the communication-optimal RAQR operating point need not coincide with the point of maximum isolated atomic sensitivity.

\subsection{Contributions and Significance}
Motivated by the above considerations, this paper develops a BER-aware communication-theoretic analysis and joint optimization framework for superheterodyne RAQR-assisted wireless communication. The principal contributions are:
\begin{enumerate}
\item We develop an RF-to-optical-to-electrical model linking four-level Rydberg dynamics, coherent optical detection, wireless multipath, and receiver noise, yielding instantaneous and average SNR and a closed-form BER for $M$-QAM as functions of RAQR physical parameters.
\item We formulate the RAQR operating-point design as a joint effective-SNR/BER constrained optimization over atomic, optical, and RF parameters.
\item We decompose the coupled design into physically meaningful parameter blocks and derive analytical update rules within a block-coordinate fixed-point framework. 
\item We evaluate the framework under low-power reception, bandwidth adaptation, detuning mismatch, fading and delay-spread conditions, outage, and coverage. 
\end{enumerate}
Results show that BER-aware joint optimization converts the physical sensitivity advantages of Rydberg sensing into measurable communication-level gains. This work establishes a direct analytical link between Rydberg-atom receiver physics and wireless communication performance by treating the RAQR operating point as an integral part of receiver design. 

\textbf{Organization:} Section~\ref{s2} shows the RAQR system and communication model. Section~\ref{s3} formulates equivalent baseband communication model. Section~\ref{s4} provides the BER-aware joint parameter optimization. Section~\ref{s5} undertakes the numerical validation followed by conclusion in Section~\ref{s6}.

\section{RAQR Receiver Model}\label{s2} 
\begin{figure}[t!]
\centering
\includegraphics[width=\linewidth]{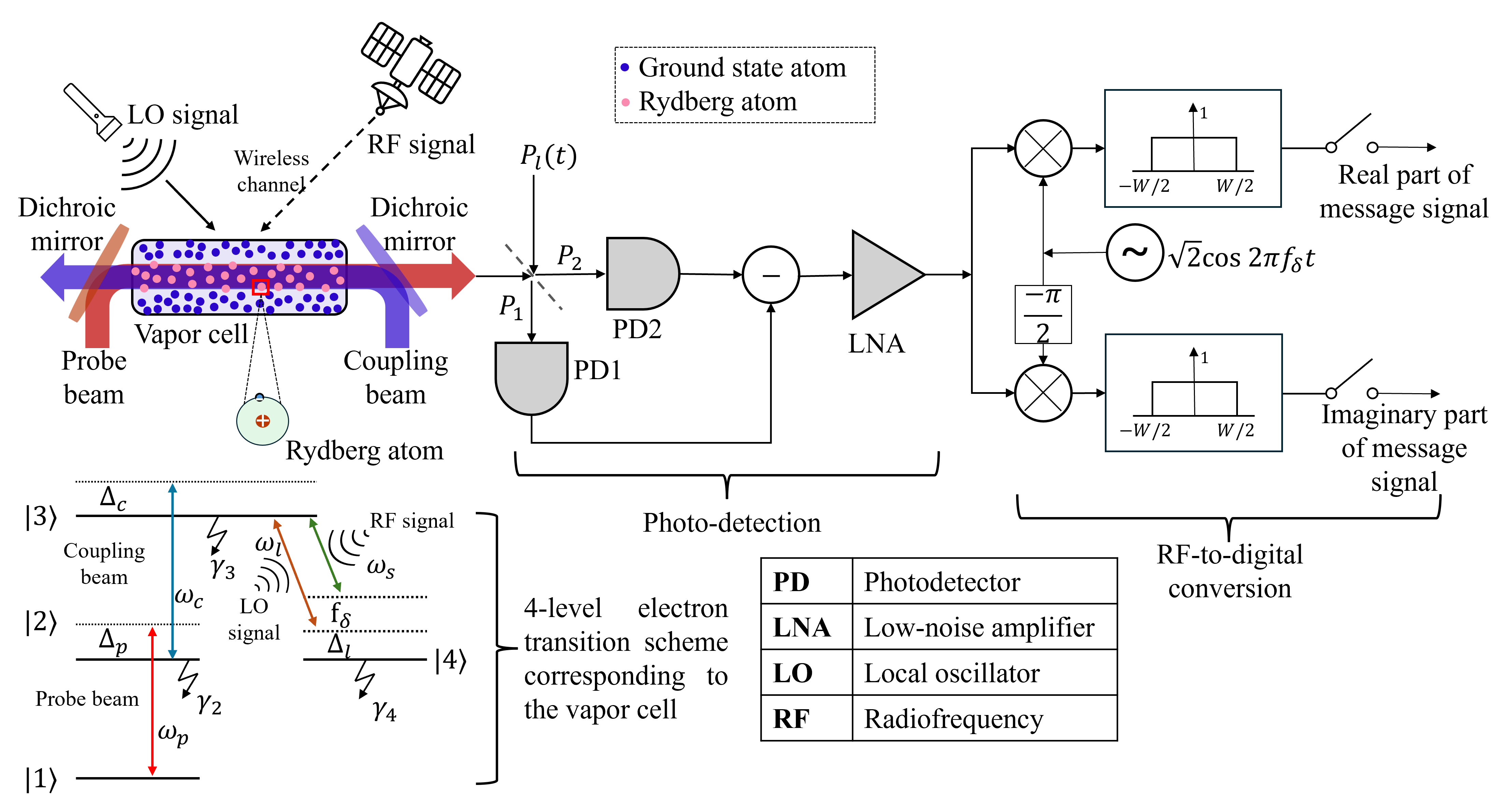}
\caption{End-to-end communication using superheterodyne Rydberg receiver involving coherent optical detection.}
\label{sysmod}
\end{figure}
The architecture of the RAQR is illustrated in Fig.~\ref{sysmod}. It comprises an alkali-atom vapor cell (e.g., Cs or Rb) and counter-propagating probe and coupling lasers that establish an EIT medium. The probe field, characterized by $\{\Omega_p,\omega_p,\Delta_p\}$, couples $|1\rangle\leftrightarrow|2\rangle$, while the coupling field $\{\Omega_c,\omega_c,\Delta_c\}$ drives $|2\rangle\leftrightarrow|3\rangle$, forming the EIT window. The incident RF field drives the adjacent Rydberg transition $|3\rangle\leftrightarrow|4\rangle$, perturbing the atomic coherence and modulating the amplitude and phase of the transmitted probe field. Photodetection of this optical response enables RF waveform recovery, thereby realizing direct RF-to-optical transduction.

In this work, we adopt the superheterodyne RAQR architecture because of its enhanced sensitivity and ability to detect arbitrarily modulated RF signals. Unlike conventional RAQRs based on incident-RF-induced Autler--Townes splitting (ATS), a LO is coherently combined with the received RF field before interaction with the atomic ensemble. Their coherent superposition produces an atomic beat response, enabling sensitive, approximately linear down-conversion of the RF signal to the optical domain. The resulting receiver can therefore be modeled as an equivalent coherent receiver, providing the basis for the subsequent communication-theoretic analysis.Consequently, the total RF Rabi frequency experienced by the atoms is given as
\begin{align}
\begin{split}
\Omega_{\rm RF}(t)=\Omega_{\rm LO}+\Omega_x(t),
\end{split}
\end{align}
where $\Omega_{\rm LO}$ and $\Omega_x(t)$ denote Rabi frequencies of the LO and received RF signal, respectively. The LO establishes the operating point of the atomic receiver, whereas the desired RF signal introduces a small perturbation about this operating point. This configuration enables linear RF-to-optical conversion with substantially improved sensitivity compared to the standard ATS configuration. The atomic dynamics are governed by the Lindblad master equation
\begin{align}
\begin{split}
\frac{d\boldsymbol{\rho}}{dt}= -\frac{i}{\hbar}[\mathbf{H},\boldsymbol{\rho}] + \mathcal{L}(\boldsymbol{\rho}),
\end{split}
\end{align}
where $\boldsymbol{\rho}$ is the density matrix, $\mathbf{H}$ is the interaction Hamiltonian, $\mathcal{L}(\cdot)$ denotes the Lindblad superoperator accounting for spontaneous emission and dephasing processes, and $[\cdot]$ is the commutator operator. The Lindblad superoperator is defined as $\mathcal{L}(\boldsymbol{\rho})=-\frac{1}{2}\{\boldsymbol{\Gamma},\boldsymbol{\rho}\}+\boldsymbol{\Lambda}$, where $\{\cdot\}$ is the anticommutator. The $\mathbf{H}$ matrix is defined as
\begin{align}
\begin{split}
\mathbf{H}=
\begin{bmatrix}
0 & \frac{\Omega_p}{2} & 0 & 0\\
\frac{\Omega_p}{2} & \Delta_p & \frac{\Omega_c}{2} & 0\\
0 & \frac{\Omega_c}{2} & \Delta_p+\Delta_c & \frac{\Omega_{\rm RF}}{2}\\
0 & 0 & \frac{\Omega_{\rm RF}}{2} & \Delta_p+\Delta_c+\Delta_{\rm LO}
\end{bmatrix}
\end{split}
\end{align}
where $\Omega_{\rm RF}$ is the Rabi frequency of the effective RF signal after the LO signal interacts with the passband received signal at RAQR. Further, $\boldsymbol{\Gamma}=\diag\{\gamma,\gamma+\gamma_2,\gamma+\gamma_3+\gamma_c,\gamma+\gamma_4\}$ and $\boldsymbol{\Lambda}=\diag\{\gamma+\gamma_2\rho_{22}+\gamma_4\rho_{44},\gamma_3\rho_{33},0,0\}$, where $\gamma_i$ is the spontaneous decay rate of the $i$th level, while $\gamma$ and $\gamma_c$ are the relaxation rates related to the atomic transition and the atomic collision effects, respectively. For tractability, we consider $\gamma=\gamma_c=0$ and assume that the decay rates of the Rydberg states $|3\rangle$ and $|4\rangle$ are comparatively low \cite{gong2026rydberg}. Then under steady-state operation, i.e., $\frac{d\boldsymbol{\rho}}{dt}=0$, the expression for $\rho_{21}$, which is associated with the probe beam as \cite{gong2026rydberg}
\begin{align}\label{rho21}
\begin{split}
&\rho_{21}(\Omega_{\rm RF})=\Omega_p\times\\
&\frac{\rho_{1r}\Omega_{\rm RF}^4+\rho_{2r}\Omega_{\rm RF}^2+\rho_{3r}-\iota(\rho_{1i}\Omega_{\rm RF}^4+\rho_{2i}\Omega_{\rm RF}^2+\rho_{3i})}{\rho_{1d}\Omega_{\rm RF}^4+\rho_{2d}\Omega_{\rm RF}^2+\rho_{3d}}
\end{split}
\end{align}
where $\rho_{r}$ and $\rho_d$ are the functions of the detuning and Rabi frequency of the coupling, probe, and LO beams \cite{gong2026rydberg}. 

\section{Equivalent Baseband Communication Model}\label{s3}
Consider a conventional RF transmitter generating a binary sequence $\mathbf{b}=\{b_1, \cdots, b_n\}$. For an $M$-QAM modulated transmission, these bits are grouped into blocks of $\log_2 M$ bits. Each bit-block is mapped to one complex constellation point $a_n\in\mathcal{A}$, where $\mathcal{A}=\{a_1, \cdots, a_M\}$ is the $M$-QAM constellation. For a normalized square $M$-QAM modulation, we have $a_n=I_n+\iota Q_n$, where $\mathbb{E}[|a_n|^2]=1$. The corresponding interpolated signal is given as
\begin{align}
\begin{split}
s_b(t)=\sqrt{P_s}\sum_{n=-\infty}^{\infty}a_ng(t-nT_s)
\end{split}
\end{align}
where $P_s$ is the average symbol power, $T_s$ is the symbol duration, and $g(t)$ is the pulse shaping filter. Assuming ideal interpolation, we have $g(t)=\sinc(Wt)$, where $W$ is the sampling rate, and $\sinc(t)=\frac{\sin(\pi t)}{\pi t}$. The upconverted passband counterpart is given  by
\begin{align}
\begin{split}
s(t)=\sqrt{2}\Re\left(s_b(t)\exp(\iota 2\pi f_x t)\right)
\end{split}
\end{align}
where $\Re(\cdot)$ is the real part operator, $\iota=\sqrt{-1}$ represents the imaginary unit, $f_x$ is the carrier frequency, and $P_s=\frac{1}{2}c\epsilon_0A_s|U_s|^2$ is the transmit power with $A_s$ denoting the effective transmitter aperture and $U_s$ being the peak electric-field amplitude corresponding to the average symbol power $P_s$. The equivalent baseband signal for one symbol duration is given as $s_b(t)=\sqrt{P_s}e^{\iota\theta_s}$. Also, it is fair to assume that $U_s$ and $\theta_s$ are time-invariant over a symbol transmission period. 

Let us consider a linear time-varying wireless communication channel $h(\tau, t)=\sum_i a_i(t)\delta[\tau-\tau_i(t)]$, with $a_i(t)$ and $\tau_i(t)$ denoting the gain and delay for the $i$th multipath component. The received basedband signal $x_b(m)$ is given as
\begin{align}
\begin{split}
x_b(t)=\sqrt{A_e}\sum_ia_i(t)\exp(-\iota 2\pi f_x\tau_i(t))s_b[t-\tau_i(t)]
\end{split}
\end{align}
where $A_e$ is the effective aperture of the RAQR. Upon sampling $x_b(t)$ at multiples of $1/W$, we obtain
\begin{align}\label{rxsignonoise}
\begin{split}
x_b(m)=\sqrt{A_e}\sum_{l}h_l(m)s_b(m-l)
\end{split}
\end{align}
where
\begin{align}\label{chnl}
\begin{split}
h_l(m)=\sum_ia_i\left(\frac{m}{W}\right)e^{-\iota 2\pi f_x\tau_i\left(\frac{m}{W}\right)}\sinc\left[l-\tau_i\left(\frac{m}{W}\right)\right]
\end{split}
\end{align}
is the equivalent discrete-time channel impulse response with $l=m-n$ being the number of discrete delay taps. The corresponding RF signal incident on the RAQR is given as
\begin{align}
\begin{split}
x(t)=\sqrt{2P_x}\cos(2\pi f_x t+\theta_x)=\sqrt{2}\Re\left\{x_b(t)e^{\iota 2\pi f_x t}\right\}
\end{split}
\end{align}
where $P_x$ and $\theta_x$ are the power and phase of the RF signal, respectively, such that $P_x=\frac{1}{2}c\epsilon_0 A_e|U_x|^2$. Considering a superheterodyne structure for the RAQR, the desired RF signal is given by $z(t)=x(t)+y_{\rm LO}(t)$, where $y(t)$ is the LO beam impinging on the RAQR, given by
\begin{align}
\begin{split}
y_{\rm LO}(t)=&\sqrt{2P_{\rm LO}}\cos(2\pi f_{\rm LO} t+\theta_{\rm LO})\\
=&\sqrt{2}\Re\left\{y_{\rm LO}^{(\rm b)}(t)e^{\iota 2\pi f_{\rm LO} t}\right\}
\end{split}
\end{align}
where $f_{\rm LO}$, $P_{\rm LO}$, and $\theta_{\rm LO}$ are the frequency, power, and phase, respectively, and $y_{\rm LO}^{(\rm b)}(t)=\sqrt{P_{\rm LO}}e^{\iota\theta_{\rm LO}}$ is the equivalent baseband signal, such that $P_{\rm LO}=\frac{1}{2}c\epsilon_0 A_e|U_{\rm LO}|^2$. Consequently, the desired RF signal $z(t)$ is given by
\begin{align}
\begin{split}
z(t)=&\sqrt{2}\Re\left\{\left[x_b(t)e^{\iota 2\pi f_{\delta}t}+y_b(t)\right]e^{\iota 2\pi f_{\rm LO} t}\right\}\\
=& \sqrt{2P_z}\cos(2\pi f_{\rm LO} t+\theta_{\rm LO})
\end{split}
\end{align}
where $f_{\delta}=f_x-f_{\rm LO}$ and $P_z=\frac{1}{2}c\epsilon_0A_e|U_z|^2$ represents the power of $z(t)$. By using the concept of vector addition, the electric field amplitude $U_z$ for $z(t)$ can be obtained as
\begin{align}
\begin{split}\label{uzform}
U_z=&\sqrt{|U_x|^2+|U_{\rm LO}|^2+2|U_x||U_{\rm LO}|\cos(2\pi f_\delta t+\theta_\delta)}\\
\stackrel{(a)}{\approx} &U_{\rm LO}+U_x\cos(2\pi f_\delta t+\theta_\delta)
\end{split}
\end{align}
where $\theta_\delta=\theta_x-\theta_{\rm LO}$ and the approximation in $(a)$ results from considering $U_x\ll U_{\rm LO}$. Further, we follow the linear relationship between the Rabi frequency and the amplitude of the RF signal as $\Omega_{\rm RF}=\frac{\mu_{34}}{\hbar}U_z$, $\Omega_{\rm LO}=\frac{\mu_{34}}{\hbar}U_{\rm LO}$, and $\Omega_x=\frac{\mu_{34}}{\hbar}U_x$, where $\mu_{34}$ represents the dipole moment corresponding to the electron transition between Rydberg states $|3\rangle$ and $|4\rangle$. Further, these substitutions in \eqref{uzform}($a$) lead to 
\begin{align}
\begin{split}
\Omega_{\rm RF}=\Omega_{\rm LO}+\Omega_x\cos(2\pi f_\delta t+\theta_\delta).
\end{split}
\end{align}
The desired RF signal interacts with the input probe beam to generate the output probe beam, which is then measured by a PD. We assume that the input laser beams have a Gaussian beam profile. Therefore, the probe beam at the access area of the atomic vapor cell is expressed as
\begin{align}\label{probepb}
\begin{split}
y_p(t)=\sqrt{2P_p}\cos(2\pi f_p t+\phi_p)=\sqrt{2}\Re\left\{y_p^{(\rm b)}e^{\iota 2\pi f_p t}\right\}
\end{split}
\end{align}
where $P_p$, $f_p$, and $\phi_p$ are the power, frequency, and phase of the input probe beam, respectively. The power can be expressed by $P_p=\frac{\pi c\epsilon_0}{8\ln 2}F_p^2|U_p|^2$, where $F_p$ represents the full width at half maximum (FWHM) of the probe beam and $U_p$ is the amplitude of the electric field corresponding to the probe beam. Furthermore, $y_p^{(\rm b)}=\sqrt{P_p}e^{\iota\phi_p}$ is the equivalent baseband signal of the passband probe beam in \eqref{probepb}. 

After propagating through the vapor cell, the probe beam is influenced by the Rydberg atoms created by interacting LO and RF beams. The amplitude $U_p'(\Omega_{\rm RF})$ and phase $\phi_p'(\Omega_{\rm RF})$ of the output probe beam (incident on the PD) is given as
\begin{align}
\begin{split}
U_p'(\Omega_{\rm RF})=&U_p\exp(-\pi d\Im\left\{\chi(\Omega_{\rm RF})\right\}/\lambda_p)\\
\phi_p'(\Omega_{\rm RF})=&\phi_p+\pi d\Re\left\{\chi(\Omega_{\rm RF})\right\}/\lambda_p
\end{split}
\end{align}
where $\Im\{\cdot\}$ is the imaginary part operator, $\lambda_p$ is the wavelength of the probe laser, $d$ is the length of the vapor cell, and $\chi(\Omega_{\rm RF})$ is the susceptibility of the atomic vapor medium. Further, the susceptibility can be given in terms of the coherence between Rydberg states $|2\rangle$ and $|1\rangle$ as
\begin{align}
\begin{split} 
\chi(\Omega_{\rm RF})=-\frac{2N_0\mu_{12}^2}{\epsilon_0\hbar\Omega_p}\rho_{21}(\Omega_{\rm RF})
\end{split}
\end{align}
where $N_0$ is the atomic density in the vapor cell and $\mu_{21}$ is the dipole moment of transition between $|1\rangle\to |2\rangle$. Thus the output probe beam $y_{\rm op}(\Omega_{\rm RF}, t)$ is given as
\begin{align}
\begin{split}
y_{\rm op}(\Omega_{\rm RF}, t)=&\sqrt{2P_{\rm op}(\Omega_{\rm RF})}\cos(2\pi f_p t+\phi_p(\Omega_{\rm RF}))\\
=&\sqrt{2}\Re\left\{y_{\rm op}^{(\rm b)}(\Omega_{\rm RF},t)e^{\iota 2\pi f_p t}\right\}
\end{split}
\end{align}
where $P_{\rm op}(\Omega_{\rm RF})=\frac{\pi c\epsilon_0}{8\ln 2}F_p^2|U_p'(\Omega_{\rm RF})|^2$ is the power of the output probe beam and the equivalent probe beam signal $y_{\rm op}^{(\rm b)}(\Omega_{\rm RF},t)$ is given as
\begin{align}
\begin{split}
y_{\rm op}^{(\rm b)}(\Omega_{\rm RF},t)=\sqrt{P_{\rm op}(\Omega_{\rm RF})}e^{\iota\phi_p}e^{\iota\frac{\pi d}{\lambda_p}\Re\{\chi(\Omega_{\rm RF}\}}
\end{split}
\end{align}

Next, we consider a balanced coherent optical detection (BCOD), wherein a strong local optical beam is used to suppress the thermal noise generated by the remaining electronic components. Let the local optical beam be expressed as
\begin{align}
\small
\begin{split}
y_l(t)=&\sqrt{2P_l}\cos(2\pi f_l t+\phi_l)=\sqrt{2}\Re\left\{y_l^{(\rm b)}(t)e^{\iota 2\pi f_l t}\right\}
\end{split}
\end{align}
where $P_l=\frac{\pi c\epsilon_0}{8\ln 2}F_p^2|U_l|^2$ with $U_l$ denoting the amplitude of the electric field of the local optical beam and $y_l^{(\rm b)}(t)=\sqrt{P_l}e^{\iota\phi_l}$ is the equivalent baseband waveform. The output probe beam and the local optical beam are combined to form two distinct optical beams, namely $y_1=\frac{1}{\sqrt{2}}[y_l(t)-y_{\rm op}(\Omega_{\rm RF}, t)]$ and $y_2=\frac{1}{\sqrt{2}}[y_l(t)+y_{\rm op}(\Omega_{\rm RF}, t)]$ which are transformed into respective output photocurrents by two photodetectors according to the relationship $I_{\rm ph}(t)=\alpha|y|^2$. The individual photocurrents are subtracted to obtain an output photocurrent as $I(t)=I_{\rm ph}^{(1)}(t)-I_{\rm ph}^{(2)}(t)$, resulting in
\begin{align}
\small
\begin{split}
I(t)=\alpha\left[y_l^{(\rm b)}(t)y_{\rm op}^{(\rm b)^{\ast}}(\Omega_{\rm RF},t)-y_l^{(\rm b)^{\ast}}(t)y_{\rm op}^{(\rm b)}(\Omega_{\rm RF},t)\right]
\end{split}
\end{align}
which follows from the principle that the optical intensity is proportional to the square of the amplitude of the incident electric field. Further $\alpha=\frac{\eta_1 q}{2\pi\hbar f_p}$ denotes the sensitivity of the photodetector, with $\eta_1$ denoting the quantum efficiency and $q$ being the charge of an electron. Considering a load resistance $R_L$ and an LNA gain of $G$, the output voltage is given as 
\begin{align}
\begin{split}
V(t)=&\sqrt{G}R_L I(t)\\
=&2\sqrt{G}R_L\alpha\sqrt{P_lP_{\rm op}(\Omega_{\rm RF})}\cos(\phi_l-\phi_p(\Omega_{\rm RF}))\\
=&2\sqrt{G}R_L\alpha\sqrt{P_lP_{\rm op}(\Omega_{\rm LO})}[\cos(\phi_l-\phi_p(\Omega_{\rm LO}))\\
&-\kappa(\Omega_{\rm LO})\cos\varphi(\Omega_{\rm LO})U_x\cos(2\pi f_\delta t+\theta_\delta)]
\end{split}
\end{align}
where $\kappa=\frac{\pi d\mu_{34}}{\lambda_p \hbar}|\chi'(\Omega_{\rm LO})|$ and $\varphi(\Omega_{\rm LO})=\phi_l-\phi'_p(\Omega_{\rm LO})+\psi_p(\Omega_{\rm LO})$ with $\chi'(\Omega_{\rm LO})$ given as 
\begin{align}
\footnotesize
\begin{split}
\Re\{\chi'(\Omega_{\rm LO})\} &= -\frac{4N_0\mu_{12}^2}{\epsilon_0\hbar}\Omega_{\rm LO}\left[\frac{2\rho_{1r}\Omega_{\rm LO}^2+\rho_{2r}}{\rho_{1d}\Omega_{\rm LO}^4+\rho_{2d}\Omega_{\rm LO}^2+\rho_{3d}}\right.\\
&\left.-\frac{(\rho_{1r}\Omega_{\rm LO}^4+\rho_{2r}\Omega_{\rm LO}^2+\rho_{3r})(2\rho_{1d}\Omega_{\rm LO}^2+\rho_{2d})}{(\rho_{1d}\Omega_{\rm LO}^4+\rho_{2d}\Omega_{\rm LO}^2+\rho_{3d})^2}\right]\\
\Im\{\chi'(\Omega_{\rm LO})\} &= \frac{4N_0\mu_{12}^2}{\epsilon_0\hbar}\Omega_{\rm LO}\left[\frac{2\rho_{1i}\Omega_{\rm LO}^2+\rho_{2i}}{\rho_{1d}\Omega_{\rm LO}^4+\rho_{2d}\Omega_{\rm LO}^2+\rho_{3d}}\right.\\
&\left.-\frac{(\rho_{1i}\Omega_{\rm LO}^4+\rho_{2i}\Omega_{\rm LO}^2+\rho_{3i})(2\rho_{1d}\Omega_{\rm LO}^2+\rho_{2d})}{(\rho_{1d}\Omega_{\rm LO}^4+\rho_{2d}\Omega_{\rm LO}^2+\rho_{3d})^2}\right].
\end{split}
\end{align}
The phase $\psi_p(\Omega_{\rm LO})$ can be written as $\psi_p(\Omega_{\rm LO})=\frac{\pi}{2}-\arctan\left(\frac{\Im\{\chi'(\Omega_{\rm LO})\}}{\Re\{\chi'(\Omega_{\rm LO})\}}\right)$. Rejecting the DC part and extracting only the time-varying component, we have:
\begin{align}
\begin{split}
\tilde{V}(t)=&2\alpha\sqrt{GP_lP_{\rm op}(\Omega_{\rm LO})}\kappa(\Omega_{\rm LO})\cos\varphi(\Omega_{\rm LO})\\
&\times U_x\cos(2\pi f_\delta t+\theta_\delta)=\sqrt{2}\Re\left\{\tilde{V}_b(t)e^{\iota 2\pi f_\delta t}\right\}.
\end{split}
\end{align}
Further, substituting $U_x=\sqrt{\frac{2P_x}{c\epsilon_0A_e}}$, we have $\tilde{V}_b(t)=\sqrt{\frac{\varrho}{A_e}}\Phi x_b(t)$, where $\varrho=\frac{4\alpha^2R_L^2G}{c\epsilon_0}P_lP_{\rm op}(\Omega_{\rm LO})\kappa^2(\Omega_{\rm LO})$ and $\Phi=\cos\varphi(\Omega_{\rm LO})e^{-\iota\theta_{\rm LO}}$. Sampling at multiples of $1/W$ as shown in Fig. \ref{sysmod}, the output is given as
\begin{align}
\begin{split}
\tilde{V}_b(m)=\sqrt{\frac{\varrho}{A_e}}\Phi x_b(m).
\end{split}
\end{align}
Using \eqref{rxsignonoise} and considering an additive noise $w(m)$, we get
\begin{align}\label{rxsig}
\begin{split}
\tilde{V}_b(m)=\sqrt{\varrho}\Phi\sum_lh_l(m)s_b(m-l)+w(m).
\end{split}
\end{align}
Since, the number of delay taps $l$ is affected by the bandwidth of the RAQR $B_{\rm RAQR}=\frac{\Gamma}{2\pi}$ (which is often termed as the EIT bandwidth/linewidth), with $\Gamma$ being the total dephasing rate of the Rydberg atoms and the maximum delay spread of the channel $\tau_{\max}=\max_{\forall i}\{\tau_i\}$, the effective number of taps are obtained as $L_{\max} \approx \lceil\min\left\{W,B_{\rm RAQR}\right\}\rceil\tau_{\max}+1$. Thus the final expression is obtained as
\begin{align}
\begin{split}
\tilde{V}_b(m)=\sqrt{\varrho}\Phi\sum_{l=0}^{L_{\max}}h_l(m)s_b(m-l)+w(m).
\end{split}
\end{align}
Further, the total receiver noise $w(m)$ is modeled as 
\begin{align}
\begin{split}
\mathbb{E}[\|w(m)\|^2]=\sigma_w^2=\frac{1}{2}\left(N_{\rm QPN}+N_{\rm PSN}+N_{\rm ITN}\right)
\end{split}
\end{align}
where $N_{\rm QPN}$, $N_{\rm PSN}$, and $N_{\rm ITN}$ are the powers of the quantum projection noise, photon shot noise, and intrinsic thermal noise, respectively. Their expressions are given as \cite{gong2026rydberg}
\begin{align}
\begin{split}
N_{\rm QPN}=&\varrho c\epsilon_0\cos^2\varphi(\Omega_{\rm LO})\frac{\hbar^2\Gamma_{\rm EIT}}{\mu_{34}^2N}B\\
N_{\rm PSN}=&q\alpha G\frac{\pi c\epsilon_0}{4\ln 2}F_p^2\left(|U_l|^2+|U_{p}'(\Omega_{\rm LO})|^2\right)B\\
N_{\rm ITN}=&k_BT_0GB
\end{split}
\end{align}
where $N=\Upsilon N_0 V$ is the number of atoms, $V$ is the atomic volume containing Rydberg atoms, $\Upsilon$ is the population rate of the Rydberg state, $K_B$ is the Boltzmann constant, and $T_0$ is the equivalent receiver noise temperature. Notably, $B$ may not be the RAQR bandwidth that we discussed earlier. Instead, it is the minimum bandwidth of the entire receiver chain affected by LNA, ADC, and other electronic components. 

\section{Performance Analysis of RAQR-Aided Communication System}\label{s4}
Based on the expressions for the received signal and the receiver noise, the instantaneous receiver SNR is given as $\gamma=\gamma_0\sum_{l=0}^{L_{\max}}|h_l|^2$, where the expression for $\gamma_0$ is given as follows
\begin{align}\label{snrexp}
\begin{split}
\gamma_0=\frac{\varrho\cos^2\varphi(\Omega_{\rm LO})P_s}{\sigma_w^2}
\end{split}
\end{align}
In writing the expression for the SINR, we have made a practical assumption that the EIT linewdith sets the limiting bandwidth for the entire receiver chain. We compute the mean $\mu_l=\mathbb{E}[h_l]$ and the covariance $R_{lm}=\mathbb{E}[h_lh_m^{\ast}]$ as
\begin{align}
\small
\begin{split}
\mu_l=&\sum_i\mathbb{E}[a_i]\mathbb{E}\left[e^{-\iota 2\pi f_x\tau_i}\sinc(l-\tau_i)\right]\\
R_{lm}=&\sum_{i}\sum_{k}\mathbb{E}\left[a_ia_k^{\ast}\frac{\sinc(l-\tau_i)\sinc(m-\tau_k)}{e^{\iota 2\pi f_x(\tau_i-\tau_k)}}\right]
\end{split}
\end{align}
Considering a Rayleigh distributed channel, we have $\mathbb{E}[a_i]=0$, $\forall i$ and $\mathbb{E}[a_ia_k^{\ast}]=\sigma_i^2\delta_{ik}$, we get
\begin{align}\label{cov}
\begin{split}
\mu_l=0 \text{ and } \sigma_{lm}^2=\sum_i\sigma_i^2\mathbb{E}[\sinc(l-\tau_i)\sinc(m-\tau_i)]
\end{split}
\end{align}
Therefore, applying the multivariate central limit theorem on \eqref{chnl}, we have $h_l\stackrel{d}{\to}\mathcal{CN}(0,\sigma_l^2)$. Repurposing the expression for $\gamma$ as $\gamma=\gamma_0\mathbf{h}^\top\mathbf{h}$, where $\mathbf{h}=[h_0, \cdots, h_{L_{\max}}]$, we get $\mathbf{h}\sim\mathcal{CN}(0,\mathbf{R})$. Further, diagonalizing $\mathbf{R}=\mathbf{U}\boldsymbol{\Lambda}\mathbf{U}^H$, we have $\mathbf{h}=\mathbf{U}\boldsymbol{\Lambda}^{1/2}\mathbf{z}$, where $\mathbf{z}\sim\mathcal{CN}(\boldsymbol{0}, \mathbf{I})$. Therefore, $\gamma=\gamma_0\sum_{k=1}^{L_{\max}+1}\lambda_k|z_k|^2$, where $|z_k|^2$ are independent unit exponential random variables and $\lambda_k$ is the $k$th Eigenvalue. We write $\gamma=\sum_k\gamma_0\lambda_kX_k$ where $X_k\sim \text{Exp}(1)$, which is exactly a generalized chi-square distribution given as
\begin{align}\label{pdfexp}
\begin{split}
f_{\Gamma}(\gamma)=\sum_{k=1}^{L_{\max}+1}\frac{C_k}{\gamma_0\lambda_k}e^{-\frac{\gamma}{\gamma_0\lambda_k}}
\end{split}
\end{align}
where $C_k=\prod_{m\neq k}\frac{\lambda_k}{\lambda_k-\lambda_m}$ which is the hypoexponential distribution. Considering a square Gray-coded $M$-QAM, the conditional BER is given as
\begin{align}\label{ber}
\begin{split}
P_b(\gamma)\approx \frac{4}{\log_2 M}\left(1-\frac{1}{\sqrt{M}}\right)Q\left(\sqrt{2\beta\gamma}\right). 
\end{split}
\end{align}
where $\beta=\frac{3}{2(M-1)}$.  Upon unconditioning the BER expression in \eqref{ber} using the density function in \eqref{pdfexp}, we get 
\begin{align}
\small
\begin{split}
\bar P_b=\frac{2}{\log_2 M}\left(1-\frac{1}{\sqrt{M}}\right)\sum_{k=1}^{L_{\max}}C_k\left(1-\sqrt{\frac{\beta\gamma_0\lambda_k}{1+\beta\gamma_0\lambda_k}}\right).
\end{split}
\end{align}
Earlier, we derived a closed-form average BER for an $M$-QAM system with an RAQR receiver, explicitly coupling channel characteristics through the covariance-matrix eigenvalues with RAQR characteristics through the equivalent gain $\Phi$, quantum noise, and finite EIT bandwidth. Thus, BER depends on both the wireless channel and RAQR operating point, motivating joint optimization of receiver and communication parameters to minimize BER under RAQR physical constraints. The BER minimization problem is given as
\begin{align}
\begin{split}
(\mathcal{P}1): \quad & \min_{\mathbf{x}}\bar P_b(\mathbf{x})\\
\mathbf{C}11: \quad & \Omega_p < \eta\Omega_c, \quad 0<\eta<\epsilon, \; \epsilon\ll 1\\
\mathbf{C}12: \quad & |\Delta_p+\Delta_c|\leq \Gamma_{\rm EIT}\\
\mathbf{C}13: \quad & |\Gamma_{\rm EIT}-2\pi W|\leq\delta, \quad \delta\ll 1\\
\mathbf{C}14: \quad & \omega_{\rm LO}=2\pi f_x\pm 2\pi f_{\delta}\\
\mathbf{C}15: \quad & -\frac{\pi}{2}\leq\varphi(\Omega_{\rm LO})\leq \frac{\pi}{2}
\end{split}
\end{align}
where $\mathbf{x}=[\Omega_p, \Omega_c, \Omega_{\rm LO}, \omega_{\rm LO}, U_{\rm LO}, U_p, U_l, \Delta_p, \Delta_c, \Delta_{\rm LO}, N_0,$ $d, F_p]^T$. $\mathbf{C}11$ meets the weak probe assumption required for the proper operation of the RAQRs, $\mathbf{C}12$ ensures the two-photon coherence condition, and $\mathbf{C}13$ ensures perfect match between the bandwidth of the transmitted signal and the EIT linewidth, which is the limiting bandwidth in the receiver chain. Notably, too narrow $\Gamma_{\rm EIT}$ decreases the noise. However, the receiver also filters out part of the transmitted signal. In contrast, a broader $\Gamma_{\rm EIT}$ leads to capturing the full signal bandwidth at the cost of more noise and inter-symbol interference. $\mathbf{C}14$ ensures the low IF operation of the RAQR and $\mathbf{C}15$ ensures a valid range output for the $\arctan$ function. 

It can be noted from $\mathbf{C}14$ that the angular frequency of the LO beam is fixed. Further, it is very clear that the objective function is only piece-wise continuous with jumps at $\Gamma_{\rm EIT}=\frac{2\pi n}{\tau_{\max}}\mathbb{1}(W\tau_{\max}\geq n)$, where $n\in \mathbb{I}$ and $\mathbb{1}(\cdot)$ is a binary indicator function. As a consequence, the objective function is not convex in $\Omega_c$. However, under the constraint $\mathbf{C}13$, we obtain $L_{\max}=W\tau_{\max}+1$. With this relaxation, the limit of the summation is independent of $\Gamma_{\rm EIT}$. Further, as each summand in the expression for $\bar P_b$ is $\geq 0$ and uncorrelated, we have $\frac{\partial\sum f}{\partial x}=\sum\frac{\partial f}{\partial x}$. Considering the argument of derivative function as $\sqrt{\frac{\beta\gamma_0\lambda_k}{1+\beta\gamma_0\lambda_k}}$, we note that maximizing $\gamma_0$ leads on to the sufficient condition for minimizing the BER. 

Next, we execute constraint $\mathbf{C}1$ through first order approximations for the terms involving $\Omega_p$. This leads to $A_i$, $B_i$ and $C_i$, $i=1,2,3$ behaving independent of $\Omega_p$. Thus, $\rho_{21}(\Omega_{\rm LO})\propto \Omega_p$, which is the multiplicative term at the start of \eqref{rho21}. With this understanding, we note that $\varrho$, $\cos\varphi$, and $|U'_p|$ render independent of $\Omega_p$. Also, note that 
\begin{align}
\begin{split}
\varphi(\Omega_{\rm LO})=\frac{\pi}{2}+\phi_l-\phi_p-\zeta(\Omega_{\rm LO})
\end{split}
\end{align}
where $\zeta(\Omega_{\rm LO})=\frac{\pi d}{\lambda_p}\Re\{\chi(\Omega_{\rm LO})\}+\arctan\left(\frac{\Im\{\chi'(\Omega_{\rm LO})\}}{\Re\{\chi'(\Omega_{\rm LO})\}}\right)$ and $\phi_l-\phi_p$ can be tuned so as to achieve $\varphi(\Omega_{\rm LO}) \approx n\pi$. With this setting we achieve $\cos\varphi(\Omega_{\rm LO})\to 1$. Also, as previously established, $U_{\rm LO}=\frac{\hbar}{\mu_{34}}\Omega_{\rm LO}$. With these simplifications and having executed constraints $\mathbf{C}11$ and $\mathbf{C}14$, the revised optimization problem is given as
\begin{align}
\small
\begin{split}
(\mathcal{P}2): \quad & \max_{\mathbf{x}'} \gamma'_0(\mathbf{x}')\quad \text{subject to} \quad \mathbf{C}12 \text{ and } \mathbf{C}13
\end{split}
\end{align}
where $\mathbf{x}'=[\Omega_p, \Omega_c, \Omega_{\rm LO}, U_p, U_l, \Delta_p, \Delta_c, \Delta_{\rm LO}, N_0, d, F_p]^T$ is the reduced set of decision variables as per the relaxed constraints and $\gamma_0=2P_s B^{-1}\gamma'_0|_{\varphi(\Omega_{\rm LO})\to n\pi}$. 

\subsection{Performance Optimization wrt $N_0$, $d$, and $F_p$}
We rewrite $\gamma'_0$ as $\gamma'_{01}$ in terms of $N_0$, $d$, and $F_p$, treating other decision variables as constants. 
\begin{align}
\begin{split}
\gamma'_{01}=\frac{A_1N_0^2d^2F_P^4e^{-2\beta d}}{B_1N_0d^2F_p^5e^{-2\beta d}+C_1F_p^2+D_1F_p^2e^{-2\beta d}+E_1}
\end{split}
\end{align}
where $A_1=c\epsilon_0G\left(\frac{\alpha R_L \pi^2 c\epsilon_0\mu_{34}U_lU_p\chi_1}{4\ln2\lambda_p\hbar}\right)^2$, $B_1=\frac{2\pi Ac\epsilon_0\hbar^2}{\mu_{34}^2\Upsilon V}$, $C_1=\frac{q\alpha G\pi c\epsilon_0}{4\ln 2}U_l^2$, $D_1=\frac{C}{U_l^2}U_p^2$, and $E_1=k_BT_0G$, such that $\chi_1=|\chi'|/N_0$ and $\beta=\frac{\pi}{\lambda_p}\Im\{\chi(\Omega_{\rm LO})\}$. Notably, $\frac{\partial\gamma'_{01}}{\partial N_0}>0$ without saturation, which means that $N_0$ must be chosen at the extreme of its limit, say $N_0=N_0^{\max}$. Next, differentiating wrt $d$ and equating to zero, we obtain
\begin{align}
\small
\begin{split}
\frac{D_1F_p^2e^{-2\beta d}+E_1+C_1F_p^2-(E_1+C_1F_p^2)\beta d}{(B_1N_0F_p^5d^2e^{-2\beta d}+D_1F_p^2e^{-2\beta d}+E_1+C_1F_p^2)^2}=0.
\end{split}
\end{align}
On solving, we get
\begin{align}
\begin{split}
d^{\ast}=\frac{1}{\beta}\left[1+\frac{1}{2}\mathcal{W}\left(\frac{2D_1F_pF_p}{E_1+C_1F_pF_p}e^{-2}\right)\right]
\end{split}
\end{align}
where $\mathcal{W}(\cdot)$ is the Lambert $W$-function. Finally, on differentiating wrt $F_p^2$ and equating to zero, we have
\begin{align}
\small
\begin{split}
\frac{B_1N_0d^2e^{-2\beta d}F_p^5-2(C_1+D_1e^{-2\beta d})F_p^2-4E_1}{(B_1N_0F_p^5d^2e^{-2\beta d}+D_1F_p^2e^{-2\beta d}+E_1+C_1F_p^2)^2}=0
\end{split}
\end{align}
which can be solved using any standard numerical technique, such as the Newton-Raphson approach. However, for analytical tractability, we analyze the above expression under two considerations: thermal noise-limited (TNL) and thermal noise-dominant (TND) regions
\begin{align}\label{fpval}
\begin{split}
F_p^{\ast}=
\begin{cases}
\sqrt[3]{\frac{2(C_1+D_1e^{-2\beta d})}{B_1N_0d^2e^{-2\beta d}}} &, \text{TNL}\\
\sqrt[5]{\frac{4E_1}{B_1N_0d^2e^{-2\beta d}}} &, \text{TND}
\end{cases}
\end{split}
\end{align}

\subsection{Performance Optimization wrt $U_p$ and $U_l$}
Rewriting $\gamma'_0$ as $\gamma'_{02}$ in terms of $U_p$ and $U_l$ we get
\begin{align}
\begin{split}
\gamma'_{02}=\frac{A_2U_p^2U_l^2}{B_2U_p^2U_l^2+C_2U_l^2+D_2U_p^2+E_1}
\end{split}
\end{align}
where $A_2=4\alpha^2R_L^2Gc\epsilon_0e^2E_p\kappa^2(\Omega_{\rm LO})$, $B_2=c\epsilon_0\hbar^2\Gamma_{\rm EIT}A_2/\mu_{34}^2N$, $C_2=0.5q\alpha Ge$, and $D_2=C_2E_p$, such that $e=\frac{\pi c\epsilon_0F_p^2}{8\ln 2}$ and $E_p=e^{-2\beta d}$. Here, we check the limit of $\gamma'_{02}$ as $U_p, U_l\to\infty$, which is essentially the limit at their respective upper bounds.
\begin{align}
\begin{split}
\lim_{U_p, U_l\to\infty}\gamma'_{02}\to \frac{A_2}{B_2}=\frac{\mu_{34}^2N}{c\epsilon_0\hbar^2\Gamma_{\rm EIT}}
\end{split}
\end{align}
We define a knee point wrt $U_l$, as the value of $U_l$  (obtained in terms of $U_p$) at which $\gamma'_{02}=\eta_1\frac{A_2}{B_2}$, leading to
\begin{align}\label{ulup}
\begin{split}
U_l=\sqrt{\frac{\eta_1(D_2U_p^2+E_1)}{(1-\eta_1)B_2U_p^2-\eta_1 C_2}}
\end{split}
\end{align}
with the feasibility condition $U_p>\sqrt{\frac{\eta_1C_2}{(1-\eta_1)B_2}}=U_p^{\min}$. We utilize the lowest possible amplitude of $U_p$ that still satisfies the relation in \eqref{ulup}. Consequently, we define a knee point again as $U_l=\eta_2\sqrt{\frac{\eta_1 D_2}{(1-\eta_1)B_2}}$, such that $\eta_2>1$, leading to
\begin{align}
\begin{split}
U_p=\sqrt{\frac{1}{\eta_2^2-1}\left[\frac{E_1}{D_2}+\frac{\eta_2^2\eta_1C_2}{(1-\eta_1)B_2}\right]}
\end{split}
\end{align}

\subsection{Performance Optimization wrt $\Delta_p$, $\Delta_c$, and $\Delta_{\rm LO}$}
Rewriting $\gamma'_0$ in terms of $\boldsymbol{\Delta}=[\Delta_p, \Delta_c, \Delta_{\rm LO}]^{\top}$, we get
\begin{align}\label{gammaDel}
\begin{split}
&\gamma'_{03}=\\
&\frac{A_3e^{-\frac{2\pi d}{\lambda_p}\Im\{\chi(\boldsymbol{\Delta})\}}|\chi'(\boldsymbol{\Delta})|^2}{B_3e^{-\frac{2\pi d}{\lambda_p}\Im\{\chi(\boldsymbol{\Delta})\}}|\chi'(\boldsymbol{\Delta})|^2+C_3e^{-\frac{2\pi d}{\lambda_p}\Im\{\chi(\boldsymbol{\Delta})\}}+D_3}
\end{split}
\end{align}
where $A_3=4\alpha^2R_L^2Gc\epsilon_0e^2U_p^2U_l^2f^2$, $B_3=\frac{c\epsilon_0\hbar^2\Gamma_{\rm EIT}}{\mu_{34}^2N}A_3$, $C_3=\frac{C_3}{U_l^2}U_p^2$, $E_3=2q\alpha GeU_l^2$, and $D_3=E_3+E_1$, such that $f=\frac{\pi d\mu_{34}}{\lambda_p\hbar}$. Therefore, we solve the following optimization
\begin{align}
\begin{split}
(\mathcal{P}3): \quad \max_{\boldsymbol{\Delta}}\gamma'_{03} \text{ subject to } \mathbf{C}12.
\end{split}
\end{align}
Differentiating the objective function wrt $\boldsymbol{\Delta}$ we get
\begin{align}
\begin{split}
\nabla_{\boldsymbol{\Delta}}\mathcal{L}_3=&\gamma'_{03}(g_{1,\gamma_{03}}\mathcal{B}_{\chi}+g_{2,\gamma_{03}}\mathcal{C}_{\chi})
\end{split}
\end{align}
where $\mathcal{D}$ denotes the denominator of \eqref{gammaDel}, $\mathcal{B}_{\chi}=\Im\{\nabla_{\boldsymbol{\Delta}}\chi(\boldsymbol{\Delta})\}$, $\mathcal{C}_{\chi}=\frac{\Re\{\chi'^{\ast}(\boldsymbol{\Delta})\nabla_{\boldsymbol{\Delta}}\chi'(\boldsymbol{\Delta})\}}{|\chi'(\boldsymbol{\Delta})|^2}$, $g_{1,\gamma_{03}}=-\frac{\pi d}{\lambda_p}\frac{D_3}{\mathcal{D}}$, and $g_{2,\gamma_{03}}= \frac{C_3\exp(-2\pi d\lambda_p^{-1}\Im\{\chi(\boldsymbol{\Delta})\})+D_3}{\mathcal{D}}$. To satisfy $\mathbf{C}12$, we use a centered parameterization, i.e., $\Delta_p=\Delta+\frac{\epsilon}{2}$, $\Delta_c=-\Delta+\frac{\epsilon}{2}$, such that $|\epsilon|\leq \Gamma_{\rm EIT}$. On simplifying
\begin{align}
\small
\begin{split}
\chi=K_{\chi}\frac{-(2\Delta+\epsilon)\Omega_{\rm LO}^4-2\Omega_c^2(\Delta_{\rm LO}+\epsilon)\Omega_{\rm LO}^2-\iota\gamma_2\Omega_{\rm LO}^4}{D}
\end{split}
\end{align}
where $K_{\chi}=\frac{2N_0\mu_{12}^2}{\epsilon_0\hbar}$ and $D=(2\Omega_p^2+\gamma_2^2)\Omega_{\rm LO}^4+2\Omega_p^2(\Omega_c^2+\Omega_p^2)\Omega_{\rm LO}^2$. Therefore, $\Im\{\chi\}=-K_{\chi}\frac{\gamma_2\Omega_{\rm LO}^4}{D}$. Notably, $\mathcal{B}_{\chi}=0$. Further, we have
\begin{align}
\begin{split}
\chi'=\frac{2K_{\chi}\Omega_{\rm LO}^5}{D^2}[H-\iota 2\gamma_2\Omega_p^2(\Omega_c^2+\Omega_p^2)]
\end{split}
\end{align}
where $H=2\Omega_c^2(2\Omega_p^2+\gamma_2^2)(\Delta_{\rm LO}+\epsilon)-2\Omega_p^2(\Omega_c^2+\Omega_p^2)(2\Delta+\epsilon)$. Therefore, the stationarity condition $\mathcal{C}_{\chi}=0$ leads to $H=0$. Therefore, we have $\Delta=\kappa\Delta_{\rm LO}+\left(\kappa-\frac{1}{2}\right)\epsilon$, where $\kappa=\frac{\Omega_c^2(2\Omega_p^2+\gamma_2^2)}{2\Omega_p^2(\Omega_c^2+\Omega_p^2)}$. This provides the optimal line rather than a unique optimal solution. Therefore, to find a unique optimal, a natural choice is to minimize the physical detuning norm. Therefore, a daughter optimization problem is formulated as
\begin{align}
\begin{split}
(\mathcal{P}3.1): \quad & \min_{\Delta}\; \|\boldsymbol{\Delta}\|^2=\Delta_p^2+\Delta_c^2+\Delta_{\rm LO}^2\\
\mathbf{C}3.11: \quad & \Delta_p=\Delta+\frac{\epsilon}{2},\; \Delta_c=-\Delta+\frac{\epsilon}{2},\; |\epsilon|\leq \Gamma_{\rm EIT}\\
\mathbf{C}3.12: \quad & \Delta=\kappa\Delta_{\rm LO}+\left(\kappa-\frac{1}{2}\right)\epsilon.
\end{split}
\end{align}
Therefore, differentiating wrt $\Delta_{\rm LO}$, we get
\begin{align}
\begin{split}
\Delta_{\rm LO}^{\ast}=-\frac{2\kappa\left(\kappa-\frac{1}{2}\right)}{1+2\kappa^2}\epsilon.
\end{split}
\end{align}
The optimal central detuning is obtained as $\Delta^{\ast}=(\kappa-0.5)(1+2\kappa^2)^{-1}\epsilon$. Thus, the optimal detuning vector is obtained as
\begin{align}\label{detUp}
\small
\begin{split}
\boldsymbol{\Delta}^{\ast}=\begin{bmatrix}
\kappa(1+\kappa)\\
1-\kappa+\kappa^2\\
2\kappa\left(\kappa-\frac{1}{2}\right)
\end{bmatrix}\frac{\epsilon}{1+2\kappa^2}.
\end{split}
\end{align}

\subsection{Performance Optimization wrt $\Omega_p$, $\Omega_c$, and $\Omega_{\rm LO}$}
Considering optimized detunings, low absorption, i.e., $\left|\frac{2\pi d}{\lambda_p}\Im\{\chi\}\right|\ll 1$, and implementing $\mathbf{C}11$, we get the rewritten SNR as function of $\boldsymbol{\Omega}=[\Omega_p, \Omega_c, \Omega_{\rm LO}]^\top$ as
\begin{align}\label{g04}
\begin{split}
\gamma'_{04}=\frac{A_4\Omega_p^4\Omega_c^4\Omega_{\rm LO}^2}{B_4\Omega_p^4\Omega_c^4\Omega_{\rm LO}^2+C_4[(2\Omega_p^2+\gamma_2^2)\Omega_{\rm LO}^2+2\Omega_p^2\Omega_c^2]^4}
\end{split}
\end{align}
where $A_4=64\alpha^2R_L^2Gc\epsilon_0\left(\frac{\pi c\epsilon_0 F_p^2}{8\ln 2}\right)^2\left(\frac{\pi d\mu_{34}}{\lambda_p\hbar}\right)^2|U_l|^2|U_p|^2K_{\chi}^2$ $\gamma_2^2$ and $B_4=A_4c\epsilon_0\frac{\hbar^2}{\mu_{34}^2N}\Gamma_{\rm EIT}$, and $C_4=q\alpha G\frac{\pi c\epsilon_0}{4\ln 2}F_p^2(|U_l|^2+|U_p|^2)+k_BT_0G$. We simplify \eqref{g04} as
\begin{align}\label{redg04}
\begin{split}
\gamma''_{04}=\frac{\Omega_p^4\Omega_c^4\Omega_{\rm LO}^2}{[(2\Omega_p^2+\gamma_2^2)\Omega_{\rm LO}^2+2\Omega_p^2\Omega_c^2]^4}.
\end{split}
\end{align}
On differentiating wrt $\Omega_p$ and setting to zero, we get
\begin{align}
\begin{split}
\Omega_p^{2\ast}=\frac{\gamma_2^2\Omega_{\rm LO}^2}{2(\Omega_c^2+\Omega_{\rm LO}^2)}.
\end{split}
\end{align}
Substituting this back in \eqref{redg04}, we note that $\frac{\partial\gamma''_{04}}{\partial\Omega_c}>0$, $\forall\Omega_c$, it is imperative that the value of SNR is maximized at the maximum budget available for $\Omega_c$. Therefore, we solve for achieving $\eta_3$ fraction of the knee point value, $\frac{1}{64\gamma_2^4\Omega_{\rm LO}^2}$. On solving, we obtain $\Omega_c^{2\ast}=\beta_c\Omega_{\rm LO}^2$, where $\beta_c=[\eta_3+\sqrt{\eta_3^2+\eta_3(1-\eta_3)}](1-\eta_3)^{-1}$. 

Re-substituting this in \eqref{redg04} we observe that $\frac{\partial \gamma''_{04}}{\partial\Omega_{\rm LO}}>0$, $\forall\Omega_{\rm LO}$. Thus, again we maximize at the limiting value  of $\Omega_{\rm LO}$, i.e., $\Omega_{\rm LO}=\Omega_{\rm LO}^{\max}$. However, from an energy conservation standpoint, we formulate a minimum norm optimization.
\begin{align}
\begin{split}
(\mathcal{P}4): \quad & \min_{\boldsymbol{\Omega}}\;\|\boldsymbol{\Omega}\|^2\\
\mathbf{C}41: \quad & \Omega_p^2=\frac{\gamma_2^2}{2(1+\beta_c)},\; \Omega_C^2=\beta_c\Omega_{\rm LO}^2, \text{ and } \mathbf{C}11.
\end{split}
\end{align}
Therefore, to satisfy the weak-probe constraint, we obtain the lower bound of the LO-Rabi frequency as
\begin{align}
\begin{split}
\Omega_{\rm LO}^{\min}=\frac{\gamma_2}{\eta\sqrt{2\beta_c(1+\beta_c)}}.
\end{split}
\end{align}
Thus, the minimum-optimum $\boldsymbol{\Omega}$ is given as
\begin{align}
\begin{split}
\boldsymbol{\Omega}^{\ast}_{\min}=\frac{\gamma_2}{\sqrt{2(1+\beta_c)}}\begin{bmatrix}
1\\
\eta^{-1}\\
(\eta\sqrt{\beta_c})^{-1}
\end{bmatrix}
\end{split}
\end{align}

\subsection{Global Assembly of the Block-Wise Optimizers}

In the preceding subsections, the optimization was carried out separately with respect to the variable blocks $(N_0,d,F_p)$, $(U_p,U_l)$, $(\Delta_p,\Delta_c,\Delta_{\rm LO})$, and $(\Omega_p,\Omega_c,\Omega_{\rm LO})$. However, these block-wise optima are mutually coupled through $\chi(\Omega_{\rm LO})$, $\chi'(\Omega_{\rm LO})$, $\Gamma_{\rm EIT}$, and the receiver gain $\varrho$. Therefore, the separately obtained closed-form solutions should be interpreted as conditional update maps rather than as independent optima. 

For $\mathbf{x}'$, the global problem can be written as
\begin{align}
\begin{split}
\mathbf{x'}^{\star}=\arg\max_{\mathbf{x}\in\mathcal{X}}\gamma_0'(\mathbf{x}),
\end{split}
\end{align}
where $\mathcal{X}$ denotes the feasible set induced by the physical and operational constraints of the RAQR, given by $\mathbf{C}11$ to $\mathbf{C}13$. Since each analytical solution is conditional on the remaining variables, we define a block-update operator $\mathcal{T}(\cdot)$ as
\begin{align}
\begin{split}
\mathcal{T}(\mathbf{x})=\left[\mathcal{T}_{\Omega},\mathcal{T}_{U},\mathcal{T}_{\Delta},\mathcal{T}_{N},\mathcal{T}_{d,F}\right]
\end{split}
\end{align}
where, a globally assembled operating point must satisfy the fixed-point condition $\mathbf{x}^{\star}=\mathcal{T}(\mathbf{x}^{\star})$. Therefore, we propose a closed-form block-coordinate fixed-point algorithm. The inner loop computes a self-consistent candidate solution for a fixed set of design parameters, while the outer loop searches over the remaining scalar parameters.

\subsubsection{Atomic density update}

From the optimization wrt $(N_0,d,F_p)$, $\gamma_0'$ is monotonically increasing in $N_0$ in the absence of saturation. Hence, the optimal density is chosen at the maximum feasible value as $N_0^{(t+1)}=N_0^{\max}$. This matches the monotonicity result obtained in the $(N_0,d,F_p)$-block.

\subsubsection{Rabi-frequency update}

If the LO upper-bound operation is imposed, then $\Omega_{\rm LO}^{(t+1)}=\Omega_{\rm LO}^{\max}$. Therefore, $\Omega_c^{(t+1)}=\sqrt{\beta_c}\Omega_{\rm LO}^{\max}$ and $\Omega_p^{(t+1)}=\frac{\gamma_2}{\sqrt{2(1+\beta_c)}}$. The weak-probe constraint is feasible if $\Omega_p^{(t+1)}<\eta\Omega_c^{(t+1)}$. Equivalently, $\eta_3$ is feasible iff
\begin{align}
\begin{split}
\Omega_{\rm LO}^{\max}>\frac{\gamma_2}{\eta\sqrt{2\beta_c(1+\beta_c)}}.
\end{split}
\end{align}
Otherwise, the current parameter tuple is discarded.

\subsubsection{EIT-bandwidth projection}

EIT linewidth is given as
\begin{align}
\begin{split}
\Gamma_{\rm EIT}^{(t+1)}=\Gamma_0+\frac{\Omega_c^{(t+1)}\Omega_c^{(t+1)}}{\gamma_2}.
\end{split}
\end{align}
The bandwidth-matching constraint is then enforced through $|\Gamma_{\rm EIT}^{(t+1)}-2\pi W|\leq\delta$. If the constraint is violated, $\Omega_c^{(t+1)}$ is projected onto the feasible interval as follows
\begin{align}
\begin{split}
\Omega_c^{(t+1)}\in\sqrt{\gamma_2}\left[\sqrt{2\pi W-\delta-\Gamma_0},\,\sqrt{2\pi W+\delta-\Gamma_0}\right].
\end{split}
\end{align}

\subsubsection{Detuning update}

For $\Omega_p^{(t+1)}$ and $\Omega_c^{(t+1)}$, compute
\begin{align}
\small
\begin{split}
\kappa^{(t+1)}=\frac{\left(\Omega_c^{(t+1)}\right)^2}{2\left(\Omega_p^{(t+1)}\right)^2}\frac{2\left(\Omega_p^{(t+1)}\right)^2+\gamma_2^2}{
\left(\Omega_c^{(t+1)}\right)^2+\left(\Omega_p^{(t+1)}\right)^2}.
\end{split}
\end{align}
The minimum-norm detuning update is undertaken as per \eqref{detUp}. This ensures $\Delta_p^{(t+1)}+\Delta_c^{(t+1)}=\epsilon$, satisfying $\mathbf{C}12$ whenever $|\epsilon|\leq\Gamma_{\rm EIT}$. The detuning update follows from the optimal line $H=0$ and the subsequent minimum-norm closure.

\subsubsection{Optical-amplitude update}

Given the current values of the other variables, recompute the constants appearing in the optical-amplitude block leading to the second knee condition
\begin{align}
\small
\begin{split}
U_p^{(t+1)}=\sqrt{\frac{1}{\eta_2^2-1}\left[\frac{E_1}{D_2}+\frac{\eta_2^2\eta_1C_2}{(1-\eta_1)B_2}\right]}.
\end{split}
\end{align}
Then the local optical amplitude is updated as
\begin{align}
\begin{split}
U_l^{(t+1)}=\eta_2\sqrt{\frac{\eta_1D_2}{(1-\eta_1)B_2}}.
\end{split}
\end{align}
Finally, the amplitudes are projected as $U_p^{(t+1)}\leftarrow\Pi_{[U_p^{\min},U_p^{\max}]}(U_p^{(t+1)})$ and $U_l^{(t+1)}\leftarrow\Pi_{[U_l^{\min},U_l^{\max}]}(U_l^{(t+1)})$.

\subsubsection{Cell-length and beam-width update}

Define $\beta^{(t+1)}=\frac{\pi}{\lambda_p}\Im\{\chi^{(t+1)}(\Omega_{\rm LO}^{(t+1)})\}$, the cell length is updated as
\begin{align}
\small
\begin{split}
d^{(t+1)}=\frac{1}{\beta^{(t+1)}}\left[1+\frac{1}{2}\mathcal{W}\left(\frac{2D_1F_p^{(t)}F_p^{(t)}e^{-2}}{E_1+C_1F_p^{(t)}F_p^{(t)}}\right)\right].
\end{split}
\end{align}
The updates are projected as $d^{(t+1)}\leftarrow\Pi_{[d^{\min},d^{\max}]}(d^{(t+1)})$ and $F_p^{(t+1)}\leftarrow\Pi_{[F_p^{\min},F_p^{\max}]}(F_p^{(t+1)})$.

\subsubsection{Convergence and global selection}

At the end of each iteration, the full reduced objective $\gamma_0'(\mathbf{x}^{(t+1)})$ is evaluated. The inner fixed-point iteration terminates when
\begin{align}
\begin{split}
\frac{\left|\gamma_0'\left(\mathbf{x}^{(t+1)}\right)-\gamma_0'\left(\mathbf{x}^{(t)}\right)\right|}{\left|\gamma_0'\left(\mathbf{x}^{(t)}\right)\right|
}<\tau.
\end{split}
\end{align}
For fixed outer-loop parameters $\boldsymbol{\theta}=[\eta_1,\eta_2,\eta_3,\epsilon,r]$, where $r\in\{\mathrm{TNL},\mathrm{TND}\}$, the converged solution is denoted by $\mathbf{x'}^{\star}(\boldsymbol{\theta})$. The globally selected operating point is obtained by
\begin{align}
\begin{split}
\mathbf{x'}_{\rm global}^{\star}=\arg\max_{\boldsymbol{\theta}\in\Theta}\gamma_0'\left(\mathbf{x'}^{\star}(\boldsymbol{\theta})\right),
\end{split}
\end{align}
where $\Theta=\left\{\eta_1,\eta_2,\eta_3,\epsilon,r:|\epsilon|\leq \Gamma_{\rm EIT},\,r\in\{\mathrm{TNL},\mathrm{TND}\}\right\}$. The corresponding objective value is $\gamma_{\rm global}^{\star}=\gamma_0'\left(\mathbf{x'}_{\rm global}^{\star}\right)$. This global assembly algorithm transforms the separately derived conditional optima into a coherent global operating point for $(\mathcal{P}2)$. The entire algorithmic flow is given in Algorithm~\ref{alg:p2}.
\begin{algorithm}[t!]
\caption{\small Global fixed-point assembly for ($\mathcal{P}2$)}\label{alg:p2}
\small
\begin{algorithmic}[1]

\REQUIRE $\eta,\eta_1,\eta_2,\eta_3,\epsilon,\Omega_{\rm LO}^{\max},W,\delta,\tau,T_{\max}$

\ENSURE $\mathbf{x}_{\rm global}^{'\star}$, $\gamma_{\rm best}$

\STATE Initialize $\gamma_{\rm best}\leftarrow -\infty$ and $\mathbf{x}_{\rm global}^{\star}\leftarrow\varnothing$

\FORALL{feasible
$\boldsymbol{\theta}
=[\eta_1,\eta_2,\eta_3,\epsilon,r]$}

    \STATE Initialize a feasible point
    $\mathbf{x}^{(0)}$

    \FOR{$t=0,1,\ldots,T_{\max}$}

        \STATE Set
        $N_0^{(t+1)}=N_0^{\max}$

        \STATE Update
        $\Omega_p^{(t+1)},
        \Omega_c^{(t+1)},
        \Omega_{\rm LO}^{(t+1)}$
        using the Rabi block

        \STATE Project
        $\Omega_c^{(t+1)}$
        to satisfy the EIT-bandwidth constraint

        \STATE Update
        $\Delta_p^{(t+1)},
        \Delta_c^{(t+1)},
        \Delta_{\rm LO}^{(t+1)}$
        using the detuning block

        \STATE Update
        $U_p^{(t+1)},
        U_l^{(t+1)}$
        using the optical-amplitude block

        \STATE Update
        $d^{(t+1)},
        F_p^{(t+1)}$
        using the cell/beam block

        \STATE Evaluate
        $\gamma_0'(\mathbf{x}^{(t+1)})$

        \IF{the convergence criterion is satisfied}
            \STATE \textbf{break}
        \ENDIF

    \ENDFOR

    \IF{$\gamma_0'(\mathbf{x}^{(t+1)})>\gamma_{\rm best}$}

        \STATE
        $\gamma_{\rm best}
        \leftarrow
        \gamma_0'(\mathbf{x}^{(t+1)})$

        \STATE
        $\mathbf{x}_{\rm global}^{'\star}
        \leftarrow
        \mathbf{x}^{(t+1)}$

    \ENDIF

\ENDFOR

\RETURN
$\mathbf{x}_{\rm global}^{\star}$,
$\gamma_{\rm best}$

\end{algorithmic}
\end{algorithm}

\section{Numerical Validation}\label{s5}
\begin{table*}[t!]
\centering
\caption{Common parameters used for numerical validation of the RAQR system.}
\label{tab:simulation_parameters}

\renewcommand{\arraystretch}{1.08}
\setlength{\tabcolsep}{4pt}

\small
\begin{tabular}{@{}l c l l | l c l l@{}}
\toprule
\multicolumn{4}{c|}{\textbf{RAQR and receiver parameters}} &
\multicolumn{4}{c}{\textbf{Communication and channel parameters}} \\
\cmidrule(lr){1-4}\cmidrule(l){5-8}

\textbf{Parameter} & \textbf{Symbol} & \textbf{Value} & \textbf{Unit} &
\textbf{Parameter} & \textbf{Symbol} & \textbf{Value} & \textbf{Unit} \\
\midrule

Cell length & $d$ & $10$ & cm & Carrier frequency & $f_c$ & $6.9458$ & GHz \\

Atomic density & $N_0$ & $4.89\times10^{10}$ & cm$^{-3}$ & Communication bandwidth & $W$ & $5$ & MHz \\

Rydberg population & $\Upsilon$ & $1$ & \% & Receiver bandwidth & $B$ & $100$ & kHz \\

Probe dipole moment & $\mu_{12}$ & $2.2327\,qa_0$ & C\,m & Modulation & $M$ & 16-QAM & -- \\

Coupling dipole moment & $\mu_{23}$ & $0.0226\,qa_0$ & C\,m & Path-loss exponent & $\beta$ & $2$ & -- \\

Rydberg dipole moment & $\mu_{34}$ & $1443.45\,qa_0$ & C\,m & Reference distance & $d_0$ & $1$ & m \\

Rydberg decay rate & $\gamma_3$ & $2\pi(3.9)$ & krad/s & Link distance & $d_{\rm link}$ & $10$--$5000$ & m \\

EIT dephasing rate & $\Gamma_2$ & $2\pi(5)$ & Mrad/s & RMS delay spread & $\tau_{\rm rms}$ & $10$--$500$ & ns \\

Probe power & $P_p$ & $20.7$ & $\mu$W & Fading margin & $M_f$ & $0,\;10$ & dB \\

Coupling power & $P_c$ & $17$ & mW & Target BER & $\overline P_b^{\,\rm tar}$ & $10^{-3}$ & -- \\

LO field amplitude & $U_{\rm LO}$ & $0.0661$ & V/m & Received RF power & $P_{\rm in}$ & $-150$ to $-60$ & dBm \\

Detector efficiency & $\eta_1$ & $0.8$ & -- & Channel model & -- & TDL/Rayleigh & -- \\

\bottomrule
\end{tabular}%
\end{table*}
Numerical validation uses a common set of atomic, optical, receiver, and communication parameters, consistent with the four-level Rydberg, coherent optical detection, and communication-theoretic models. The principal parameters are summarized in Table~\ref{tab:simulation_parameters}, while experiment-specific settings, including coupling-power sweep, detuning mismatch, delay spread, fading margin, and communication distance, are given in the corresponding subsections.
\subsection{Feasibility and Operating-Regime Characterization}
\begin{figure}[t!]
\centering
\subfigure[]{\includegraphics[width=1.13in]{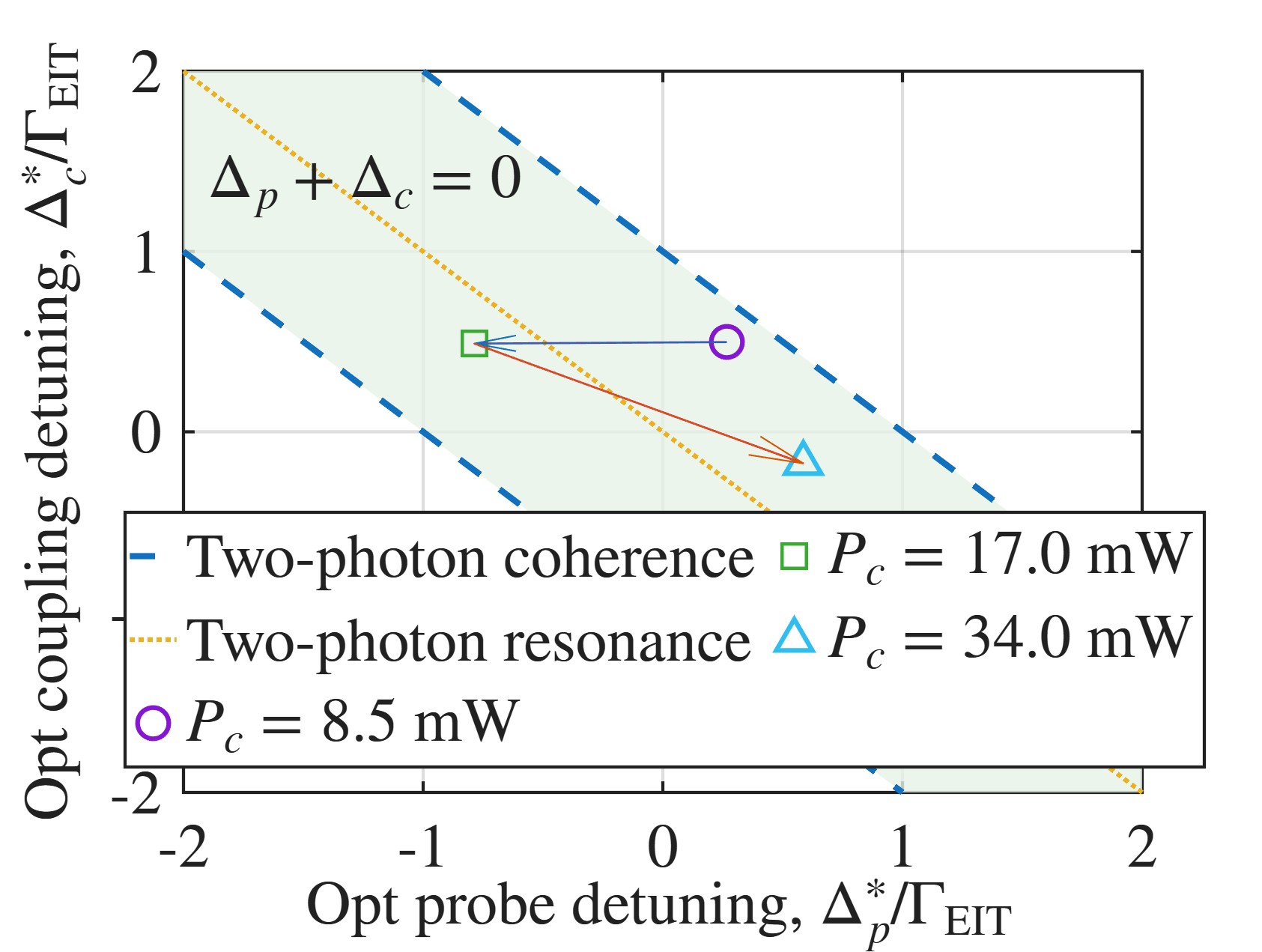}}
\subfigure[]{\includegraphics[width=1.13in]{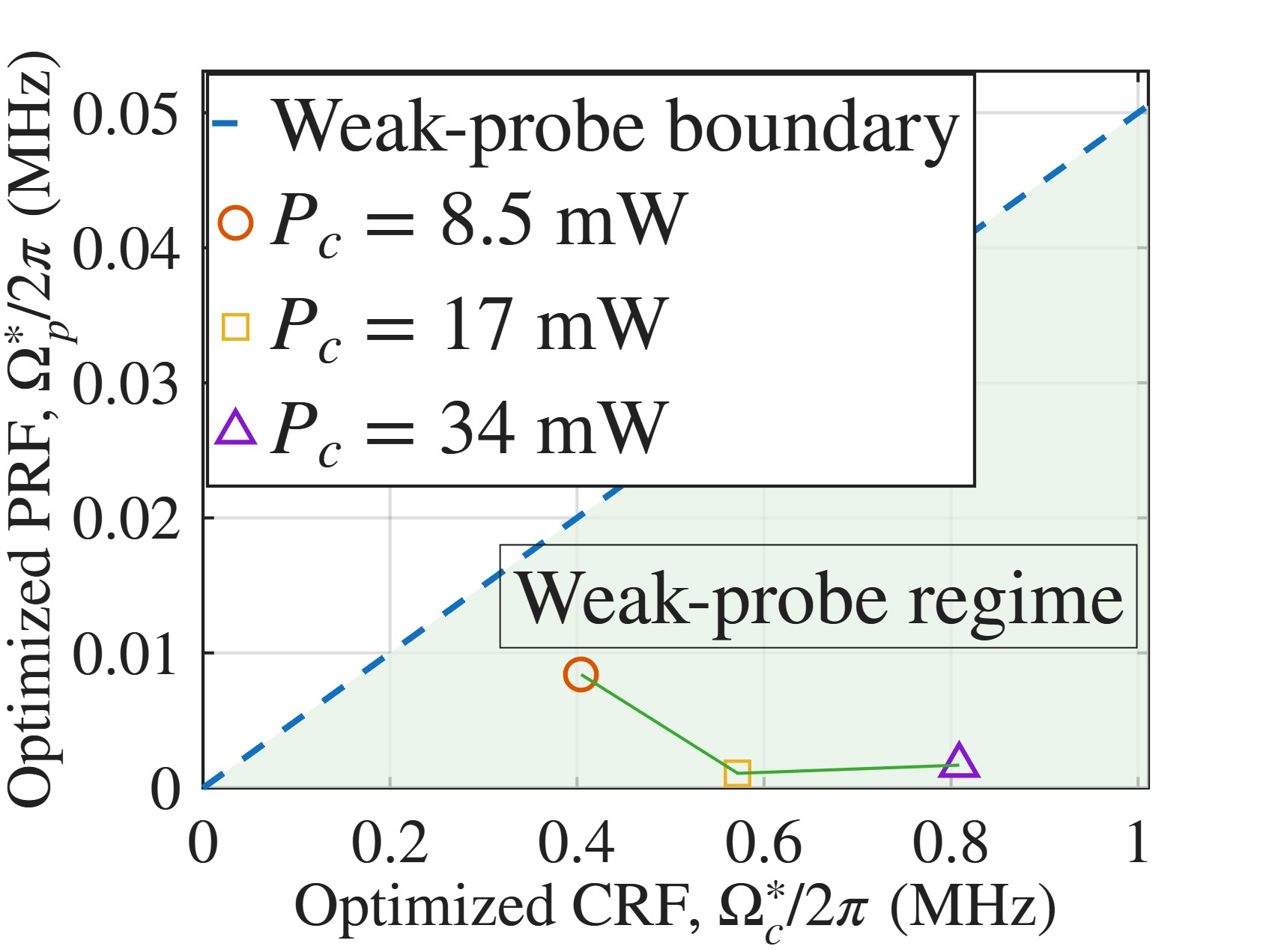}}
\subfigure[]{\includegraphics[width=1.13in]{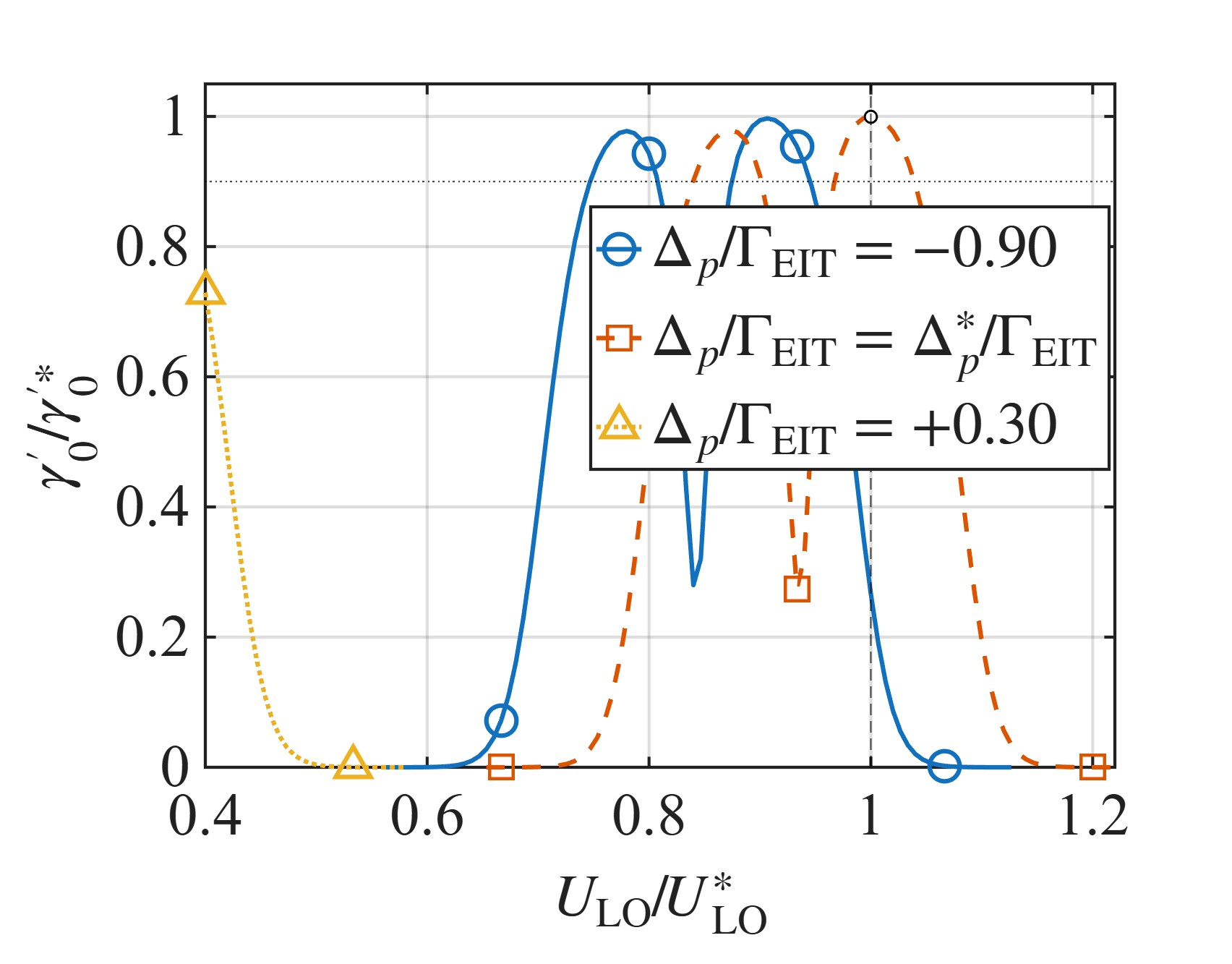}}
\caption{(a) Optimized probe and coupling detunings for different coupling powers under two-photon coherence constraint, (b) optimized probe and coupling Rabi frequencies for different coupling-power conditions, and (c) normalized RF-response sensitivity versus RF local-oscillator amplitude for representative two-photon detuning conditions.}
\label{fig1}
\end{figure}
Fig.~\ref{fig1} establishes the feasibility of the operating points used in the subsequent studies. In Fig.~\ref{fig1}(a), the optimized probe and coupling fields remain within the admissible weak-probe regime, confirming to the assumptions of the adopted EIT-based RAQR model. In Fig.~\ref{fig1}(b), the optimized operating points satisfy the two-photon detuning constraint, while varying with receiver operating conditions. This confirms that the communication-optimal point generally requires jointly optimizing the optical detunings rather than fixing them independently of communication conditions. Fig.~\ref{fig1}(c) demonstrates compatibility between the EIT response bandwidth and signal bandwidth. The optimized linewidth balances noise suppression against signal filtering and multipath-induced distortion: a narrow response reduces admitted noise but can attenuate the desired waveform, whereas an excessively broad response increases noise bandwidth. Thus, the feasible operating region represents a joint sensitivity--bandwidth tradeoff rather than simple linewidth maximization. 

\subsection{Receiver Sensitivity \& Joint Operating-Point Optimization}
\begin{figure}[t!]
\centering
\subfigure[]{\includegraphics[width=1.15in]{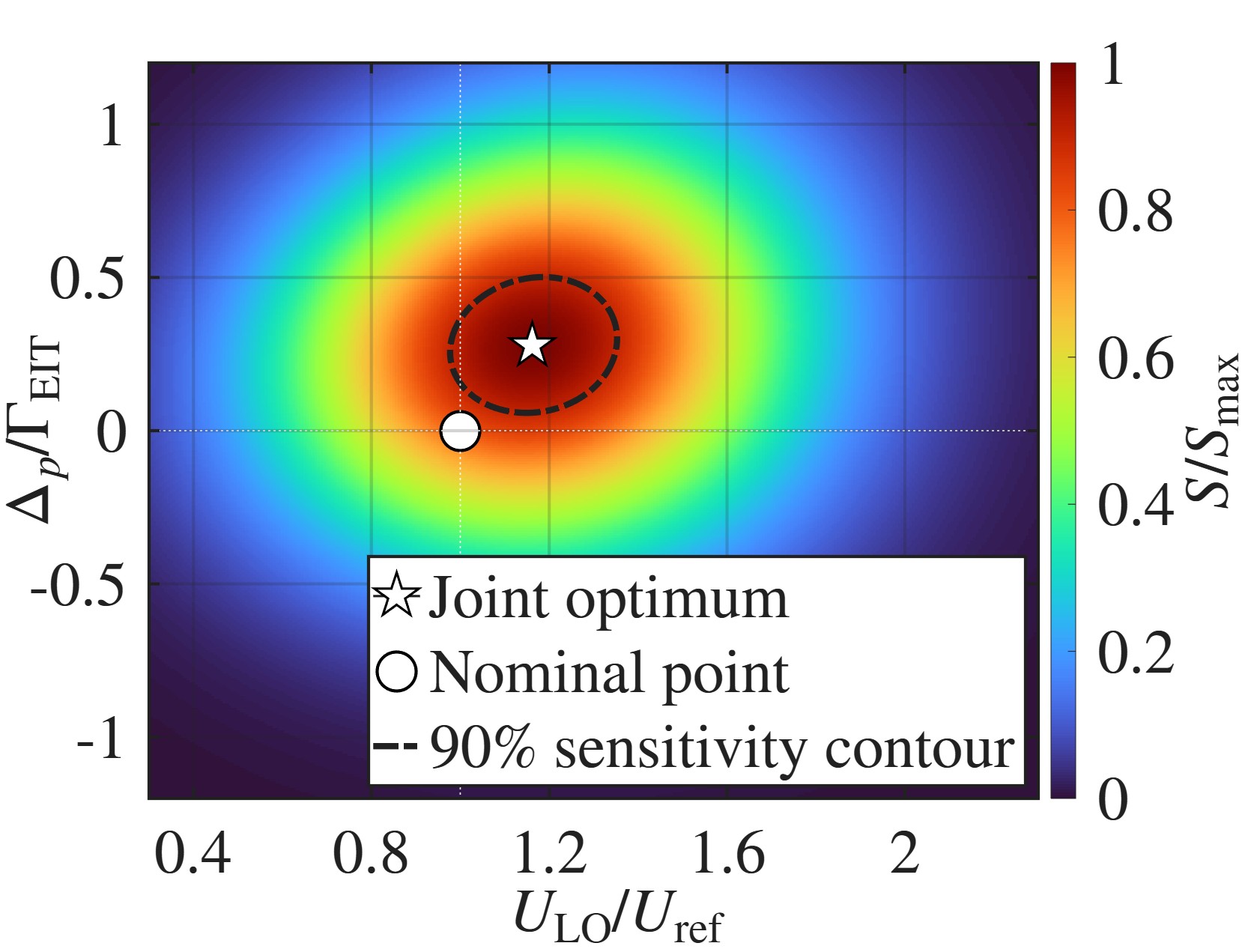}}
\subfigure[]{\includegraphics[width=1.15in]{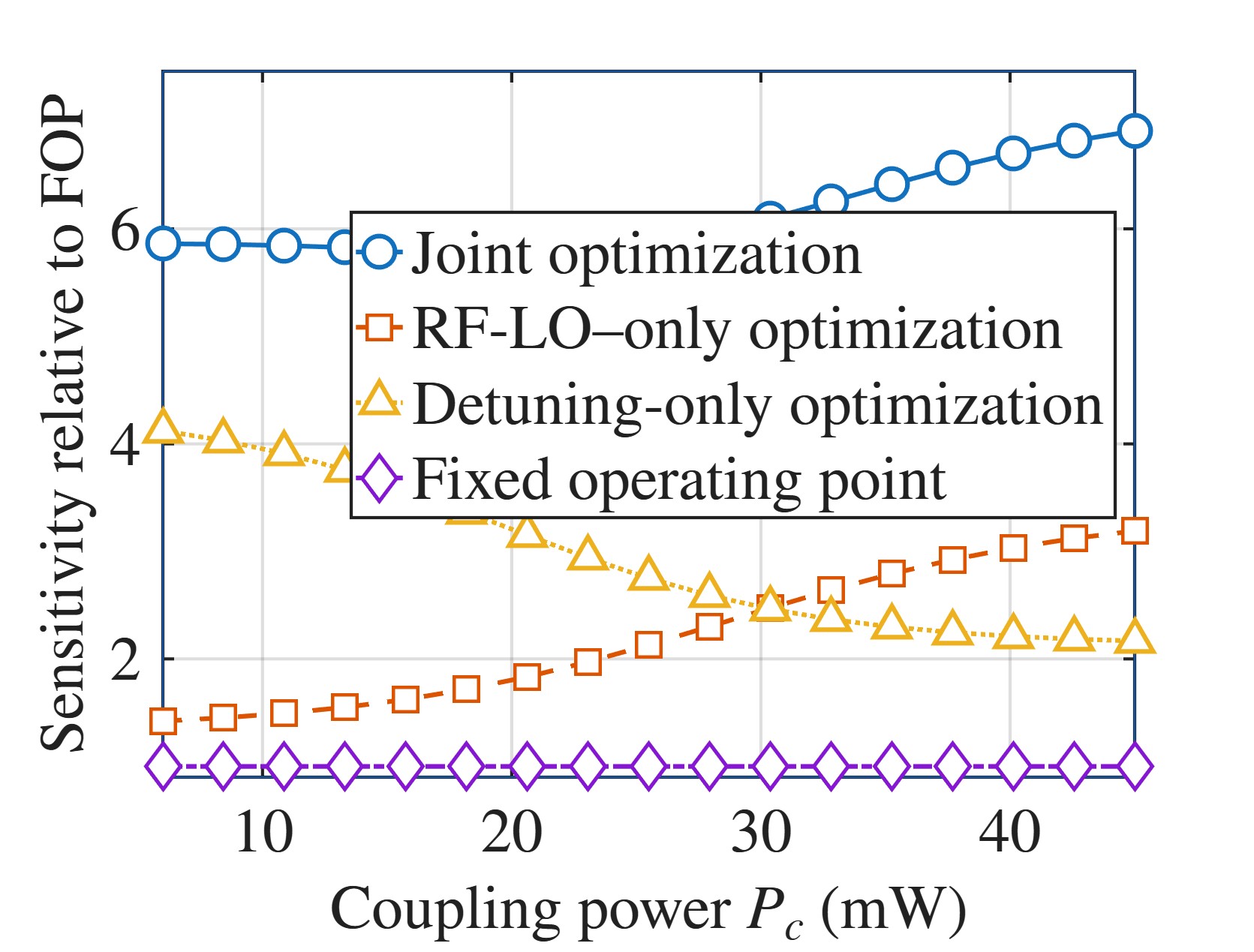}}
\subfigure[]{\includegraphics[width=1.1in]{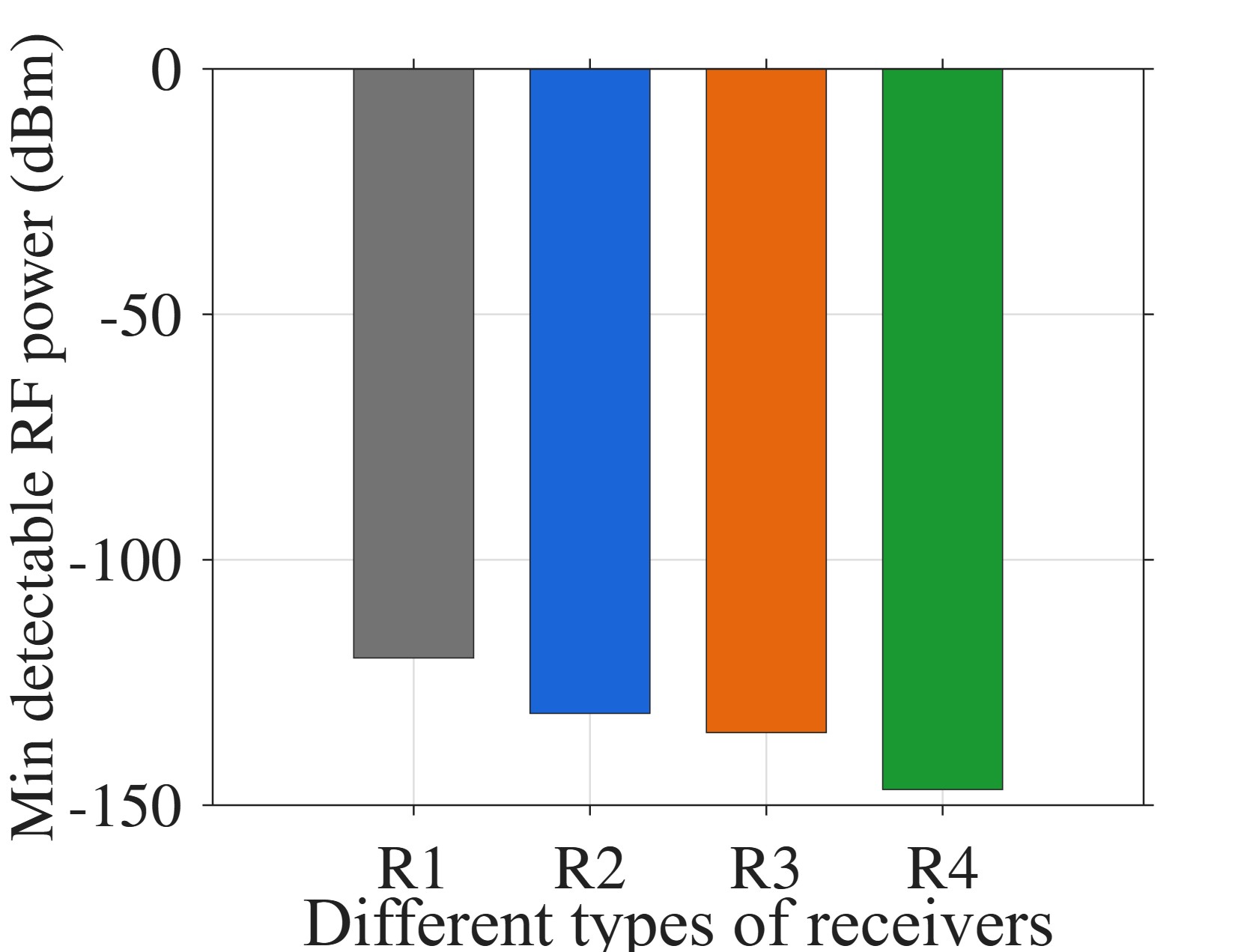}}
\caption{(a) Normalized communication sensitivity versus RF-LO amplitude and probe detuning, (b) sensitivity relative to the fixed operating point (FOP) versus coupling power, and (c) minimum detectable received RF power for conventional, advanced, fixed-RAQR, and jointly optimized RAQR receivers.}
\label{fig2}
\end{figure}
Fig.~\ref{fig2} evaluates the sensitivity enhancement from RAQR operating-point optimization and its resulting communication benefit. Fig.~\ref{fig2}(a) shows the joint sensitivity landscape versus RF local-oscillator amplitude and probe detuning, revealing a distinct high-sensitivity region and confirming that the receiver response depends jointly on the RF-LO operating point and optical detuning. Thus, independent parameter selection does not necessarily yield the communication-optimal operating point. By exploiting their coupled effects on the atomic response and RF readout, joint optimization achieves a stronger response than the nominal operating point.

Fig.~\ref{fig2}(b) compares joint optimization with restricted strategies. The jointly optimized RAQR consistently achieves the highest sensitivity, while optimization of only the RF-LO amplitude or optical detunings provides smaller gains. This confirms that the enhancement arises from the coupled effects of RF-LO interaction, optical detunings, EIT response, and receiver noise rather than from a single tunable parameter.

Fig.~\ref{fig2}(c) benchmarks the four receiver configurations, requiring approximately $-112.0$, $-125.5$, $-132.1$, and $-148.4$~dBm for the conventional RF receiver, advanced RF receiver, fixed RAQR, and jointly optimized RAQR, respectively. Thus, the advanced receiver improves sensitivity by $13.5$~dB, while joint optimization provides a further a $26.4$~dB reduction in required incident RF power relative to the conventional receiver, demonstrating that its receiver-level advantage stems not only from the Rydberg front end but also from optimizing its physical operating point.

\subsection{Robustness to Two-Photon Detuning Mismatch}
\begin{figure}[t!]
\centering
\subfigure[]{\includegraphics[width=1.13in]{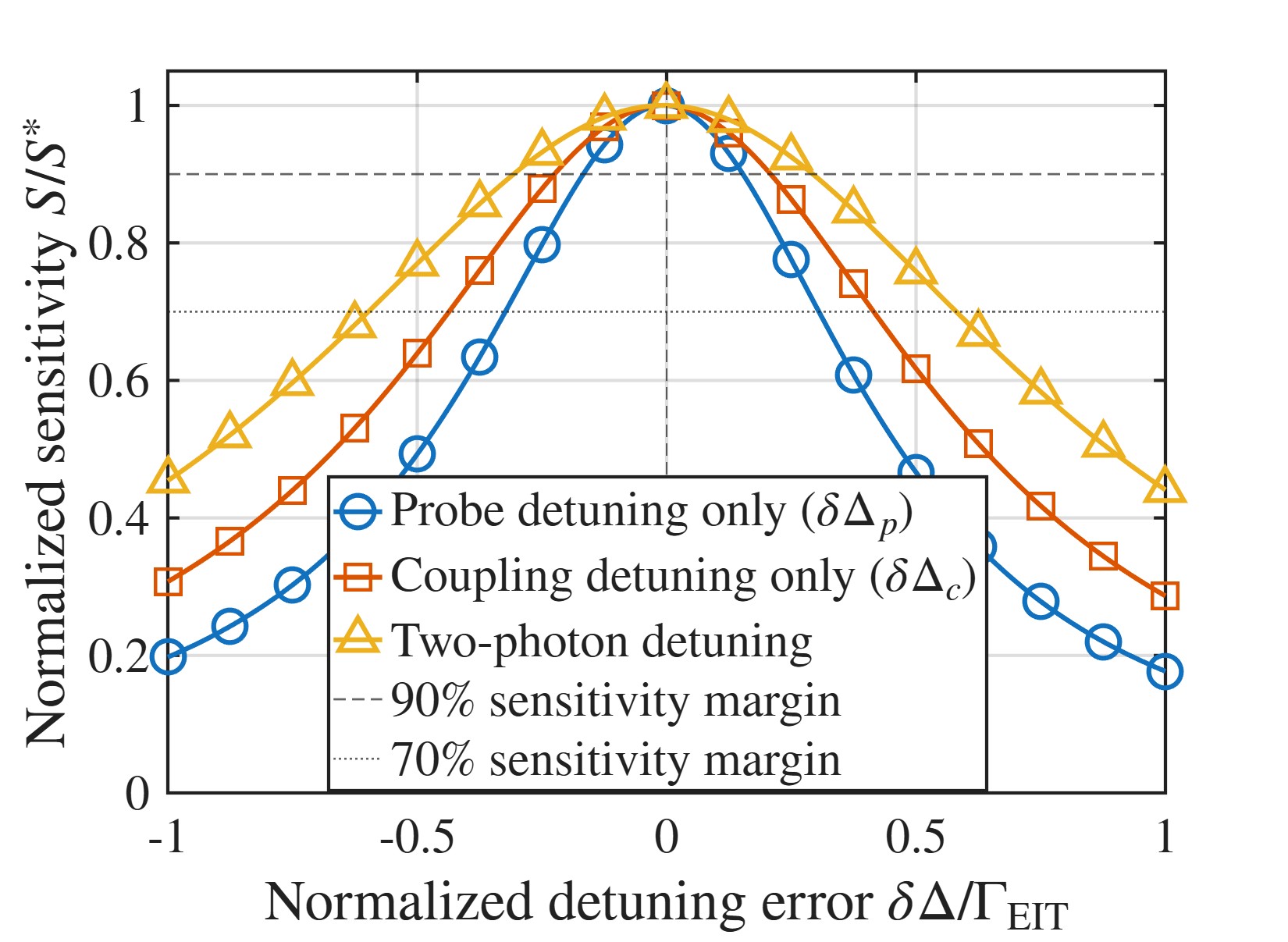}}
\subfigure[]{\includegraphics[width=1.13in]{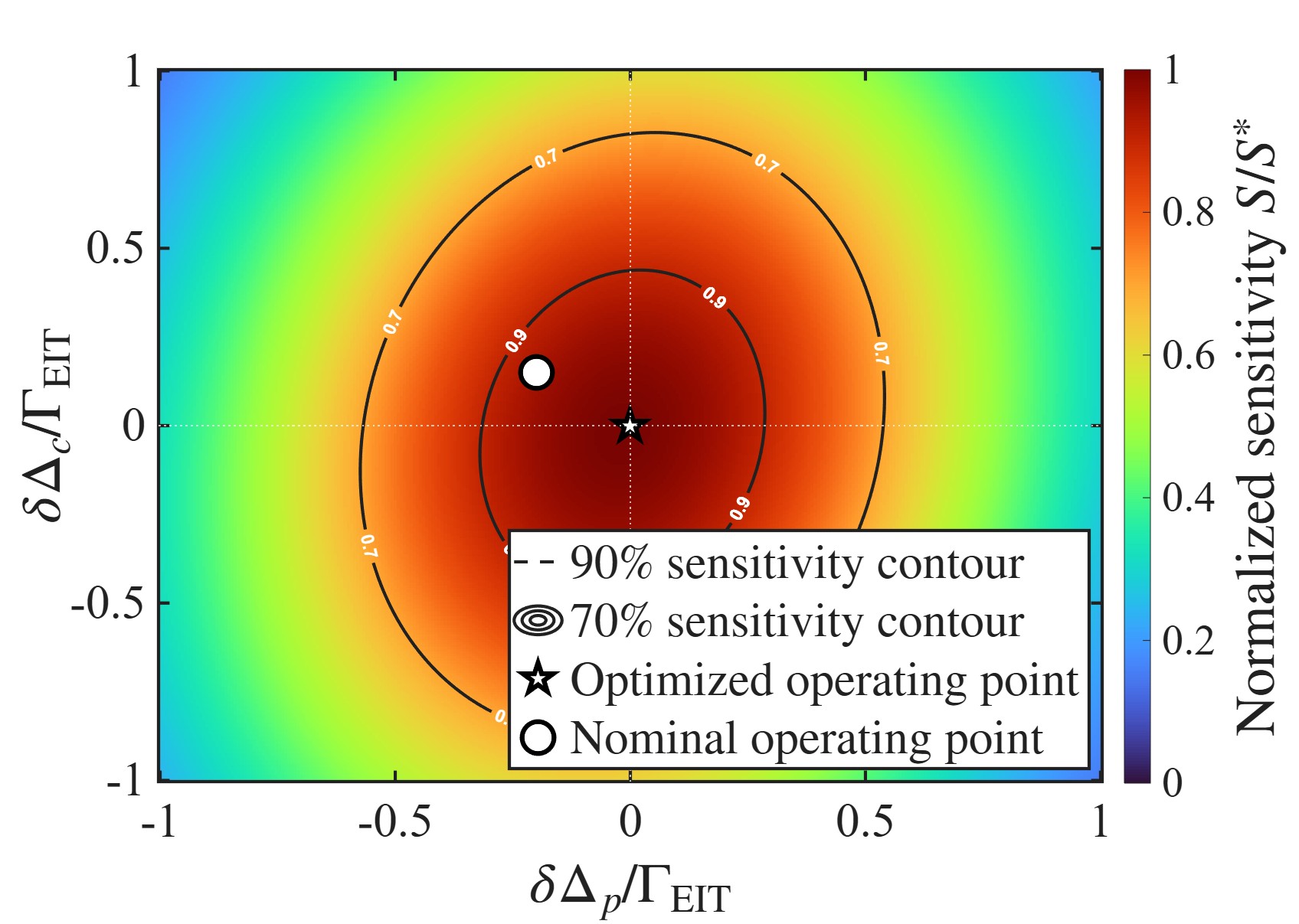}}
\subfigure[]{\includegraphics[width=1.13in]{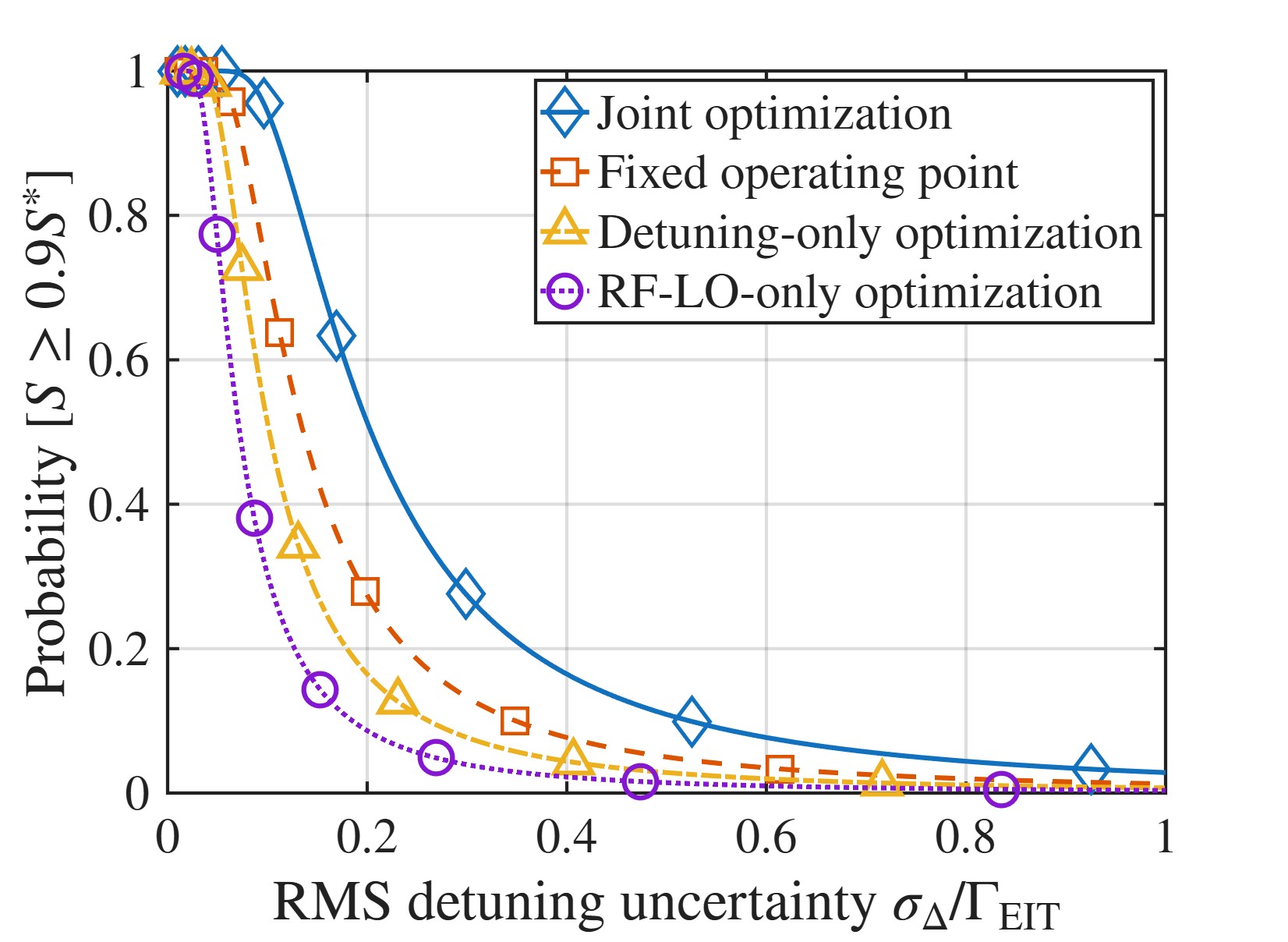}}
\caption{(a) Normalized sensitivity vs. normalized detuning error, (b) normalized sensitivity under two-photon detuning mismatch, and (c) probability of retaining at least $95\%$ of the optimized sensitivity vs. RMS detuning uncertainty.}
\label{fig3}
\end{figure}
Fig.~\ref{fig3} evaluates RAQR sensitivity to deviations from its optimized operating point, since practical utility requires adequate performance under calibration errors and temporal drift. Fig.~\ref{fig3}(a) shows that probe detuning causes greater sensitivity degradation than coupling-detuning mismatch, while common-mode (two-photon) detuning has a broader tolerance region. This asymmetric response arises from the distinct roles of the probe and coupling fields in establishing EIT susceptibility, highlighting the importance of accurate probe-frequency control. Fig.~\ref{fig3}(b) confirms this behavior for simultaneous probe- and coupling-detuning errors, yielding an anisotropic high-sensitivity region around the optimum rather than equal tolerance along both axes. The $1$-dB and $3$-dB contours therefore quantify the practically usable operating region and required optical tuning accuracy. Fig.~\ref{fig3}(c) shows that, at the $90\%$ sensitivity-retention criterion, the jointly optimized RAQR tolerates an RMS normalized detuning uncertainty of approximately $0.24\Gamma_{\mathrm{EIT}}$, versus $0.10\Gamma_{\mathrm{EIT}}$ for the restricted operating point, corresponding to an approximately $2.4$ times larger robustness radius. Thus, joint optimization enhances not only peak sensitivity but also tolerance to operating-point drift.

\subsection{BER Performance and Required Incident RF Power}
\begin{figure}[t!]
\centering
\subfigure[]{\includegraphics[width=1.71in]{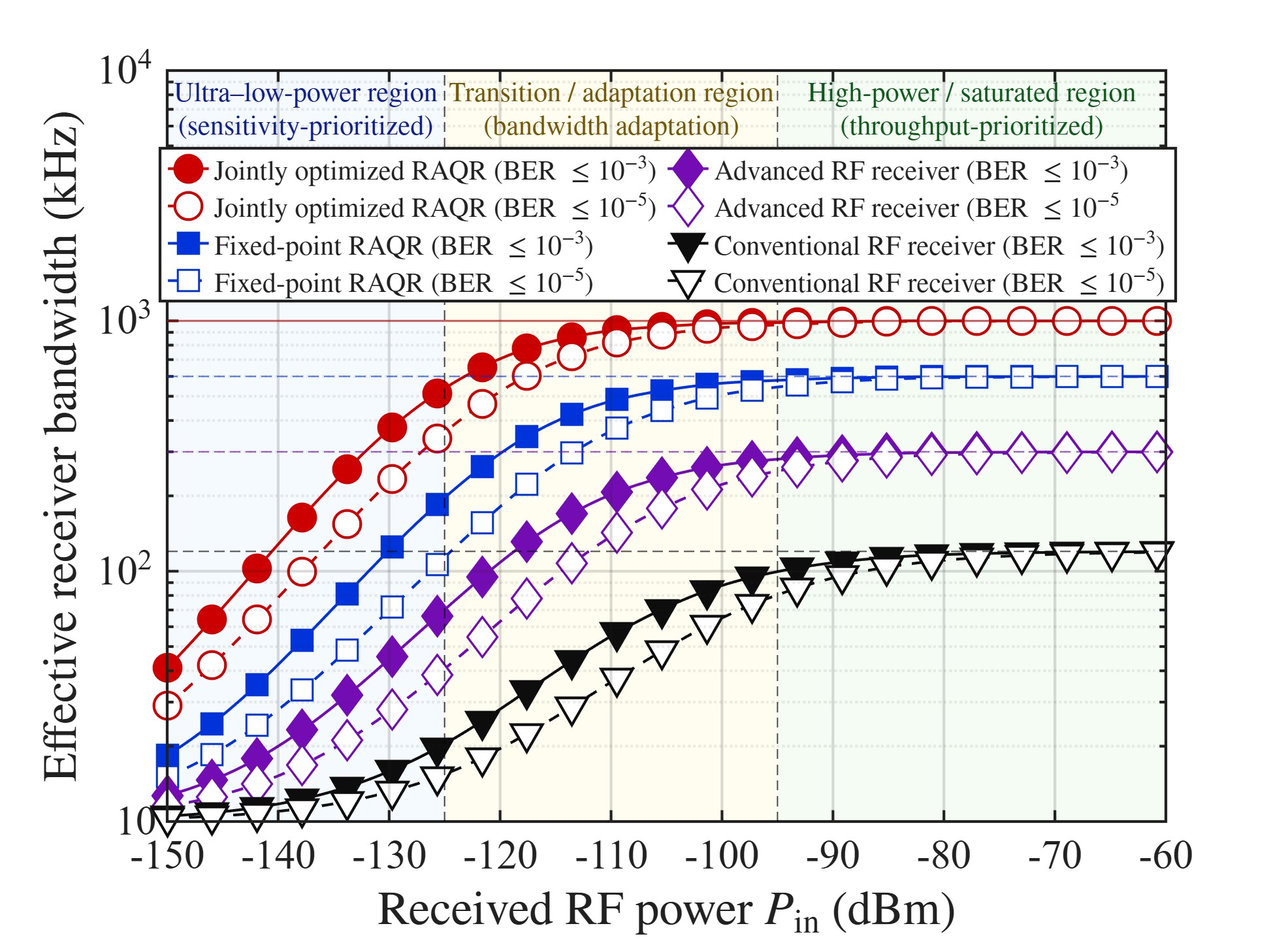}}
\subfigure[]{\includegraphics[width=1.71in]{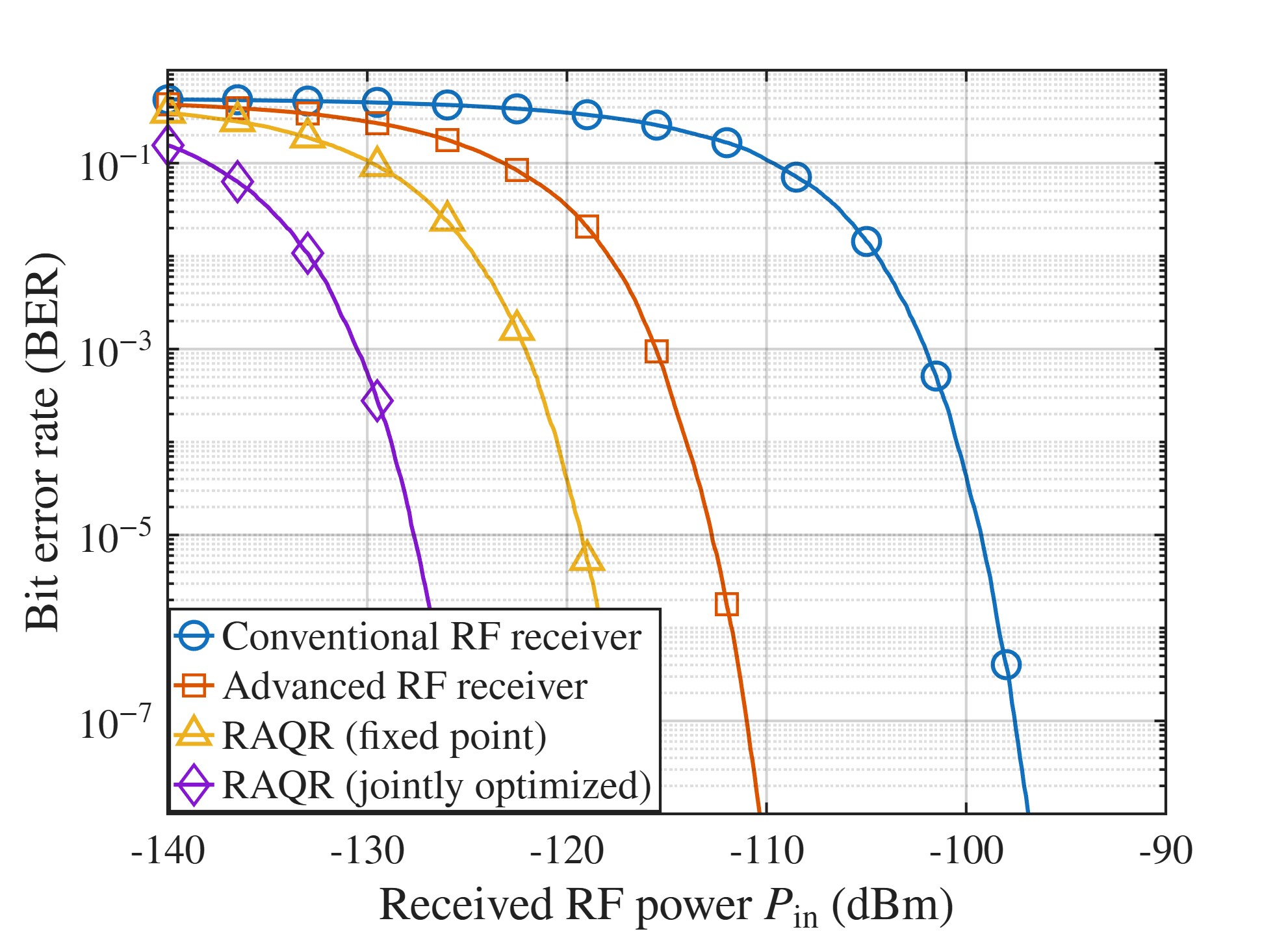}}
\caption{(a) Effective receiver bandwidth as a function of received RF power and (b) average BER versus received RF power under coherent 16-QAM reception for conventional, advanced, fixed-RAQR, and jointly optimized RAQR.}
\label{fig4}
\end{figure}
Fig.~\ref{fig4} evaluates the communication-level gains through effective bandwidth and BER. In Fig.~\ref{fig4}(a), the jointly optimized RAQR provides the most favorable adaptation under both $10^{-3}$ and $10^{-5}$ BER requirements, restricting bandwidth at ultra-low power for detectability and expanding it as power increases. For $10^{-3}$ BER, it reaches $\sim0.9$--$1.0$~MHz versus $0.6$, $0.25$, and $0.1$~MHz for the fixed-point RAQR, advanced, and conventional receivers, respectively. At $P_{\mathrm{in}}=-130$~dBm, it still supports $\sim250$~kHz versus $\sim100$~kHz for the fixed-point RAQR and only a few tens of kHz for the advanced receiver; the same ordering holds for $10^{-5}$ BER, albeit at higher power. Thus, joint optimization enables a transition from sensitivity- to throughput-prioritized operation.

Fig.~\ref{fig4}(b) shows coherent $16$-QAM BER performance. At $10^{-3}$ BER, the conventional, advanced, fixed-point, and jointly optimized receivers require approximately $-102$, $-115.5$, $-122$, and $-130.4$~dBm, corresponding to sensitivity gains of $13.5$, $20.1$, and $28.4$~dB over the conventional receiver and an additional $8.4$~dB over the fixed-point RAQR. The advantage persists at $10^{-5}$ BER, although higher power is required. Overall, joint optimization enables reliable reception at substantially lower RF power while retaining bandwidth expansion with available link margin.

\subsection{Outage and Coverage Enhancement}
\begin{figure}[t!]
\centering
\subfigure[]{\includegraphics[width=1.71in]{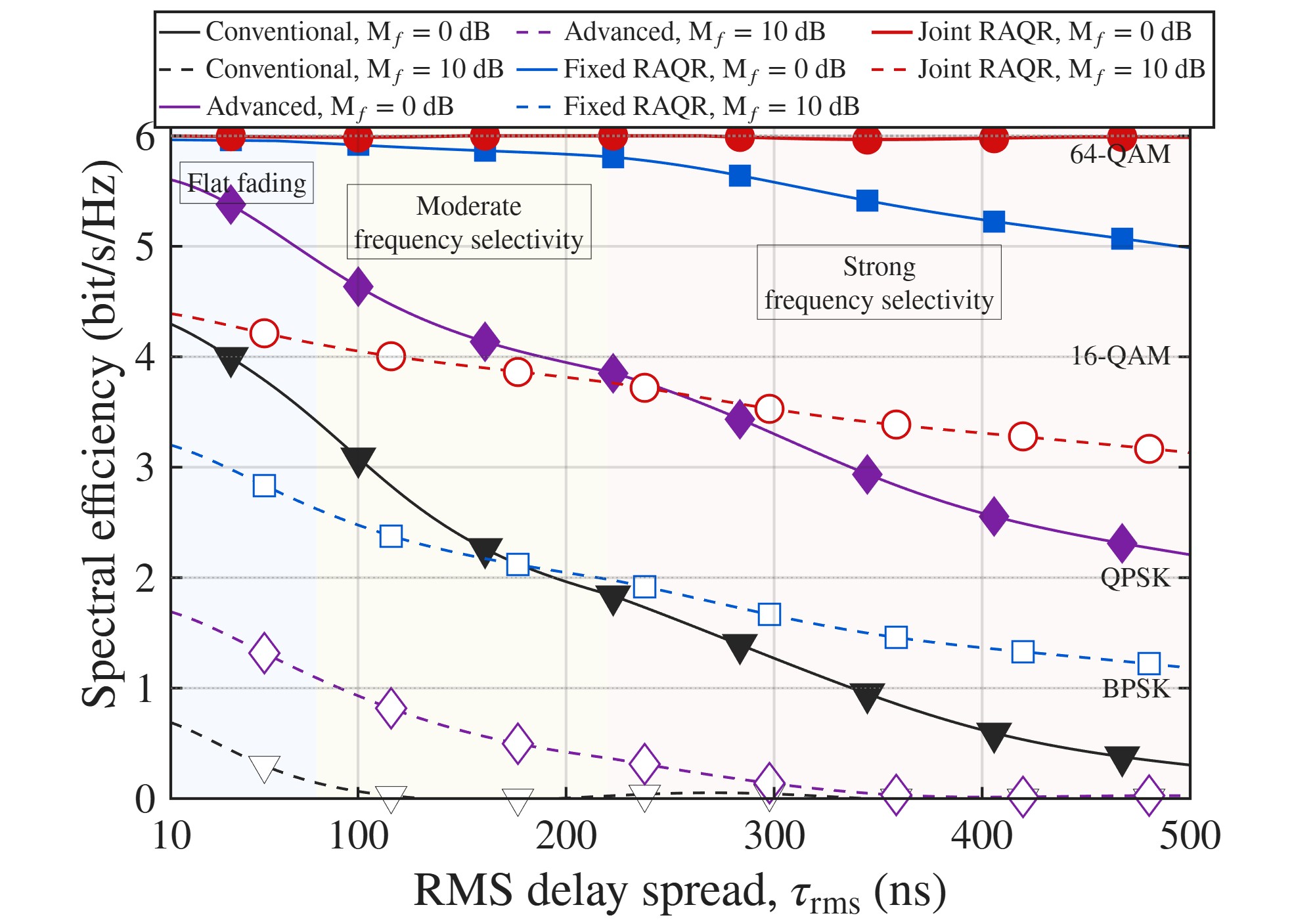}}
\subfigure[]{\includegraphics[width=1.71in]{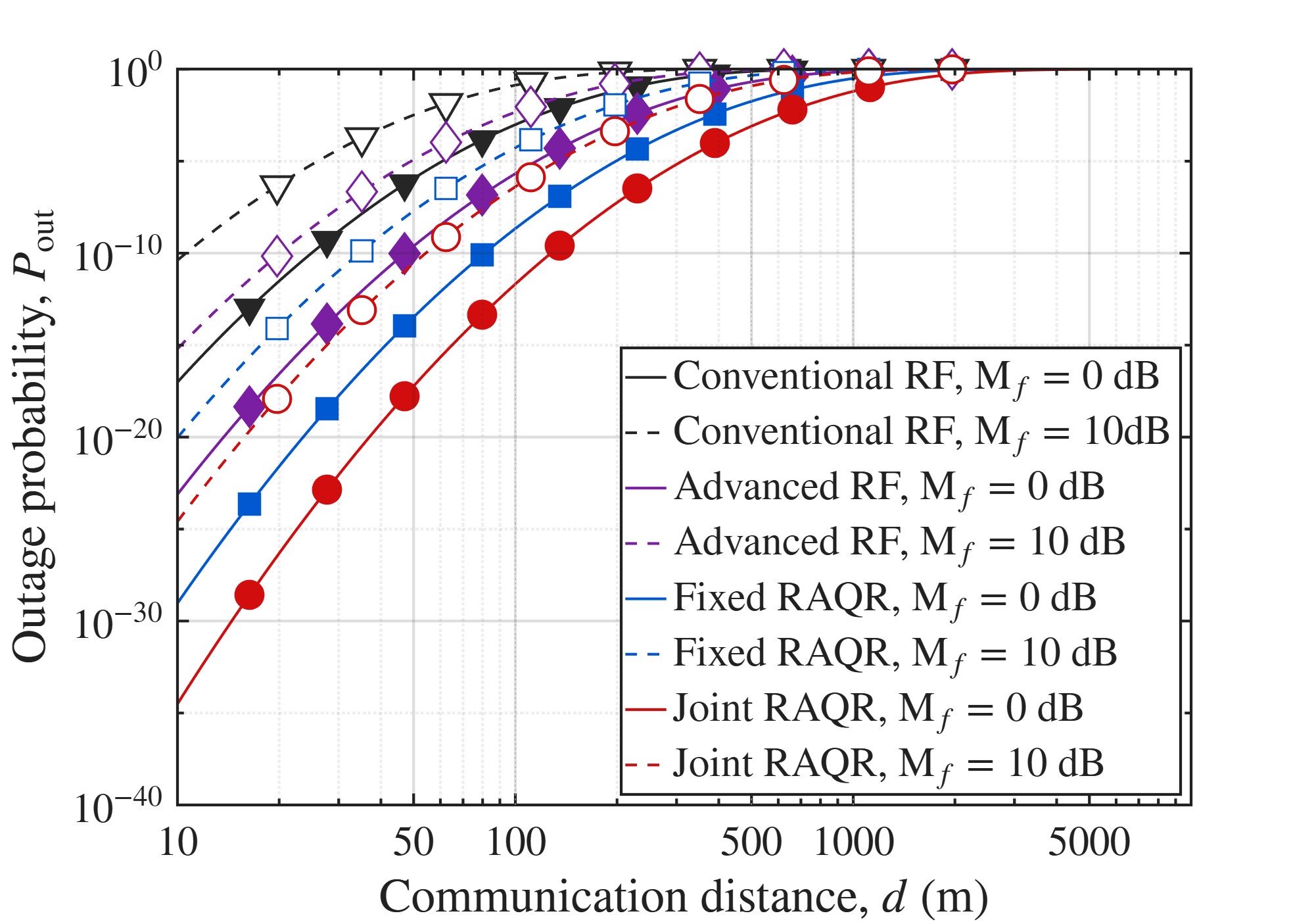}}
\caption{(a) Achievable spectral efficiency versus RMS delay spread and (b) outage probability versus communication distance for conventional, advanced, fixed-RAQR, and jointly optimized RAQR receivers under two fading-margin conditions.}
\label{fig5}
\end{figure}
The receiver-level sensitivity gain is further evaluated under frequency-selective fading and distance-dependent loss. Fig.~\ref{fig5}(a) shows spectral efficiency versus RMS delay spread, $\tau_{\mathrm{rms}}$, for two fading margins. At $\tau_{\mathrm{rms}}\approx500$~ns, the conventional, advanced, fixed-point RAQR, and jointly optimized RAQR achieve approximately $0.3$, $2.2$, $5$, and $6$~bit/s/Hz, respectively, with the jointly optimized RAQR retaining nearly the full $64$-QAM efficiency for $M_f=0$~dB. For $M_f=10$~dB, spectral efficiency decreases for all receivers, but the jointly optimized RAQR still achieves $\sim3.2$~bit/s/Hz versus $\sim1.2$~bit/s/Hz for the fixed-point RAQR and nearly zero for the conventional and advanced receivers. Thus, the joint-optimization gain persists under strong frequency selectivity and conservative fading margins.

Fig.~\ref{fig5}(b) shows outage probability versus communication distance. For both $M_f=0$ and $10$~dB, outage increases monotonically with distance as propagation loss reduces the received RF power. The jointly optimized RAQR exhibits the lowest outage over the useful range, followed by the fixed-point RAQR, advanced RF, and conventional RF receivers, consistent with the sensitivity hierarchy in Fig.~\ref{fig4}(b). The $10$-dB margin shifts all reliable ranges toward shorter distances, but the jointly optimized RAQR retains the largest useful range. Thus, the proposed receiver provides greater reliable communication coverage even at a lower received RF power.

\section{Conclusion}\label{s6}
This paper developed a BER-aware communication-theoretic framework for superheterodyne RAQR-assisted wireless communication. An end-to-end RF-to-optical-to-electrical model was established, yielding closed-form SNR and BER characterization over correlated Rayleigh fading. The RAQR operating point was jointly optimized over atomic, optical, RF, and geometric parameters using analytical block-wise updates within a fixed-point procedure while enforcing the relevant physical and bandwidth constraints. Numerical results demonstrate that the communication-optimal RAQR operation can provide substantial gains beyond isolated atomic sensitivity optimization, including improved low-power reception, adaptive bandwidth utilization, detuning robustness, resilience to fading and delay spread, lower outage, and extended communication coverage. These results establish the importance of BER-aware quantum-receiver design rather than treating the RAQR a fixed front end. The proposed framework provides a foundation for future extensions to adaptive modulation, more general fading channels, multi-antenna architectures, experimentally calibrated non-idealities, and real-time BER-aware operating-point adaptation.

\ifCLASSOPTIONcaptionsoff
\fi
\bibliography{IEEEabrv,IEEEref}{}
\bibliographystyle{IEEEtran}
\nocite{*}

\end{document}